\documentclass[twocolumn]{aastex631}
\usepackage{lineno}

\accepted{for publication in ApJ, May 15, 2023}

\begin{document}

\title{Magnetic Activity-Rotation-Age-Mass Relations in Late Pre-main Sequence Stars}

\correspondingauthor{Konstantin Getman}
\email{kug1@psu.edu}

\author[0000-0002-6137-8280]{Konstantin V. Getman}
\affiliation{Department of Astronomy \& Astrophysics \\
Pennsylvania State University \\ 
525 Davey Laboratory \\
University Park, PA 16802, USA}

\author[0000-0002-5077-6734]{Eric D. Feigelson}
\affiliation{Department of Astronomy \& Astrophysics \\
Pennsylvania State University \\ 
525 Davey Laboratory \\
University Park, PA 16802, USA}

\author[0000-0002-7371-5416]{Gordon P. Garmire}
\affiliation{Huntingdon Institute for X-ray Astronomy\\
LLC, 10677 Franks Road\\
Huntingdon, PA 16652, USA}

\begin{abstract}
We study the four-dimensional relationships between magnetic activity, rotation, mass and age for solar-type stars in the age range $5-25$~Myr.  This is the late-pre-main sequence (l-PMS) evolutionary phase when rapid changes in star’s interior may lead to the changes in magnetic dynamo mechanisms. We carefully derive rotational periods and spot sizes for 471  members of several l-PMS open clusters using photometric light curves from the Zwicky Transient Facility. Magnetic activity was measured in our previous {\it Chandra}-based study, and additional rotational data were obtained from other work. Several results emerge. Mass-dependent evolution of rotation through the l-PMS phase agrees with astrophysical models of stellar angular momentum changes, although the data point to a subpopulation of stars with slower initial rotations than commonly assumed.  There is a hint of the onset of unsaturated tachoclinal dependency of X-ray activity on rotation, as reported by Argiroffi et al. (2016), but this result is not confidently  confirmed. Both X-ray luminosity and star spot area decrease approximately as $t^{-1}$ for solar mass stars suggesting that spot magnetic fields are roughly constant and l-PMS stars follow the universal solar-scaling law between the X-ray luminosity and surface magnetic flux. Assuming convective dynamos are dominant, theoretical magnetic fluxes fail to reveal the universal law for l-PMS stars that enter late Henyey tracks. Altogether we emerge with a few lines of evidence suggesting that the transition from the turbulent to solar-type dynamo occurs at the later stages of l-PMS evolution as stars approach the Zero-Age Main Sequence.
\end{abstract}

\keywords{Pre-main sequence stars (1290) --- Stellar rotation (1629) --- Single x-ray stars (1461) --- X-ray stars (1823) ---- Protoplanetary disks (1300) --- Stellar magnetic fields (1610)}

\section{Introduction} \label{sec:intro}

For main sequence solar-type stars, the relationships between stellar magnetic activity, rotation and age has been studied for decades \citep{Skumanich72} and their origins are well-understood \citep{Schrijver00}.  Stars have a complicated rotational history during their early stages of evolution \citep{Bouvier07} but, once settled on the zero-age main sequence (ZAMS), they undergo a slow, sometimes punctuated spin-down due to the torque of a magnetized wind.  This leads to a slow weakening of their stellar activity: X-ray flares, ultraviolet emission, chromospheric spectroscopic indicators, and other manifestations of magnetic fields generated in a $\alpha\Omega$ tachoclinal dynamo.  An example of this behavior is the relation between X-ray emission and Rossby number showing saturated and unsaturated dynamo regimes for $>40$~Myr old stars in main sequence open clusters \citep{Wright11}.  

Skumanich-type relationships are not present during the pre-main sequence (PMS) phase.  No relationship between X-ray emission and rotation is seen during the `early-PMS' (henceforth e-PMS) phase  when the stars are fully convective descending the Hayashi tracks and rotation is regulated by coupling to the accreting disk \citep{Alexander2012, Henderson2012, Rebull2018}. Instead, X-ray activity is linked to bulk properties such as stellar mass, surface area, and volume \citep{Preibisch05,Getman22}.  The intense X-ray luminosities, frequent superflares, large star spots, and strong surface fields point to a highly efficient distributed $\alpha^2$ magnetic dynamo expected in fully convective stars \citep{Chabrier06, Dobler2006, Kapyla08, Christensen2009, Browning2008, Brown11, Yadav15}. In the solar-type $\alpha\Omega$ dynamo, a relatively weak magnetic field is generated in the thin tachocline at the interface between radiative and convective zones. Higher rotation rates lead to higher surface magnetic fields. Calculations of $\alpha^2$ dynamos in fully convective stars indicate strong fields are efficiently generated throughout the stellar interior reaching equipartition strengths. The energy of convective flows rather than rotation sets the strength of surface magnetic field.

The transition between the e-PMS and ZAMS phases probably occurs during the `late-PMS' (l-PMS) phase that we define to start when the star lies at the bottom of the Hayashi track on the Hertsprung-Russell diagram and to end when it begins hydrogen burning on the ZAMS\footnote{
Here we assume 'classical' PMS evolutionary tracks as initially calculated by \citet{Iben65} and more recently by many other researchers \citep{Squicciarini22}. There are substantial differences among the tracks and the evolution may further differ when the effects of accretion, internal magnetic fields, and surface star spots are considered \citep{Baraffe10, Feiden13, Somers15, Kunitomo17}.
}.
Little is known observationally about l-PMS surface magnetic activity or changes in internal magnetic dynamo processes.   For a solar mass star, the surface temperature becomes hotter while the luminosity increases and radius slowly decreases so that it moves quickly across the Hertzprung-Russell diagram.  Internally, the radiative zone grows, the convective zone size and convective turnover timescale falls, and brief nuclear burning of $^{12}$C may occur \citep{Iben65, Landin22}.  $ROSAT$ X-ray data show that stars in the IC~2391, IC~2602 and NGC~2547 open clusters exhibit the same X-ray-rotation relation as seen on the ZAMS  \citep{Patten96, Stauffer97, Irwin08,Wright11}. This suggests a transition in dynamo processes is largely complete at ages $\sim 40-50$~Myr. 

Observational evidence for the transition from the e-PMS convective dynamo to the main sequence tachoclinal dynamo is sparse.   The most tantalizing study is based on the $Chandra$ observatory detection of $\sim 400$ members of the h~Per cluster with age $\sim 13$~Myr \citep[][henceforth A16]{Argiroffi2016}.  While most stars lie in the (super)saturated regimes of the X-ray-Rossby number diagram similar to e-PMS stars, several slowly rotating stars with periods $P \gtrsim 3$~days exhibit X-ray emission declining with Rossby number suggesting that a tachoclinal dynamo has emerged.  This result is best seen in their $1.2$~M$_\odot$~$< M < 1.4$~M$_\odot$ mass stratum.   

In the present study and its earlier companion \citep[][henceforth Paper~I]{Getman22}, we seek to improve the observational evidence of l-PMS changes in interior structure and dynamos on activity-rotation-age relations.  Paper~I reports X-ray observations of 10 open clusters in the age range $5-25$~Myr at distances 0.3 to 2.5~kpc, including h~Per (= NGC 869) studied by A16.  With $Gaia$ observations to assist with selecting cluster members and $Chandra$ observations to measure magnetic activity in the X-ray band, Paper~I identifies $6,003$ stars with masses above $0.4-1.0$~M$_\odot$ completeness limits.  Most of these stars lie in the l-PMS evolutionary phase.  For comparison with fully convective e-PMS stars, these samples are combined with 40,041 stars with ages $<$5~Myr from previous $Chandra$ surveys \citep{Feigelson13, Getman2019}. 

Paper~I found that X-ray luminosities are high and constant during the first few Myr after which they decay.  The activity of higher mass stars that evolve more rapidly through the l-PMS phase falls more rapidly than in lower mass stars.  This decay in X-ray activity is interpreted as decreasing efficiency of the $\alpha^2$ dynamo as radiative cores grow. But the emergence of an $\alpha\Omega$ dynamo was not clearly seen. 

Several observational challenges must be tackled for a successful understanding of magnetic activity and dynamos in the l-PMS phase:
\begin{enumerate}

    \item  The use of $L_X/L_{bol}$ as a measure of X-ray activity that is effective for revealing dynamo activity for ZAMS and older stars cannot be effectiely applied to l-PMS stars: the changes in X-ray emission are often overwhelmed by the radical changes in bolometric luminosity and stellar radius as stars move across the HR diagram (Paper I).  We use $L_X$ alone to quantify surface magnetic activity in the l-PMS era. 
    
    \item Changes in magnetic activity during the l-PMS should be studied with a range of masses as higher mass stars evolve from the e-PMS to ZAMS regimes much more rapidly than lower mass stars.  A range of cluster ages must be examined to capture the l-PMS phase in stars of different masses.  While Skumanich-type relations are three dimensional on the main sequence, they involve four dimensions during PMS evolution: activity, rotation, age and mass.
    
    \item The X-ray brightnesses of an ensemble of stars in a single age and mass stratum exhibit considerable scatter due, at least in part, to frequent powerful superflares during the PMS phase \citep{Getman2021}. For stars not caught during flares, it is important to incorporate X-ray nondetections to give unbiased evaluation of magnetic activity using `survival' statistical methods such as the Kaplan-Meier estimator \citep{Feigelson85}.  
    
    \item Unlike e-PMS stars that are concentrated in dense clusters associated with their natal molecular environment, l-PMS stars are often dispersed.  The closest sample at distance $\sim 120$~pc, the Sco-Cen Association with outliers like the TW Hya Association and $\beta$ Pic Moving Group, is spread over large regions of the celestial sphere.  For this reason, the l-PMS clusters in Paper~I were selected to be compact. Most of these clusters are at distances around $1-2.5$~kpc.  The greater distance, however, reduces the fraction of X-ray detections with realistic $Chandra$ exposure times. 
      
    \item  The measurement of stellar rotation from photometric light curves can be difficult, particularly if a single telescope is used so the star is observable only for a few hours each day.  Alias structure in the periodogram can lead to ambiguities in the inferred rotation period \citep{VanderPlas2018, Rebull2022}.  Careful interpretation of periodograms is necessary. 
    
\end{enumerate}

We tackle these challenges in Paper~I and the current study.  Items 2$-$5 require that the magnetic activity of large stellar samples with a range of masses and ages be examined as in Paper I.  An additional $\sim 2000$ cluster members are found in the outskirts of these clusters with reduced {\it Chandra} sensitivity (see Table~4 in Paper~I). The X-ray nondetections are reported based on cluster membership obtained from $Gaia$ satellite astrometry and photometry.  Most of these stars are in the l-PMS phase and, when combined with $\sim 40,000$ stars in the e-PMS phase from earlier studies \citep{Feigelson13, Getman17}, large portions of the activity-age-mass space is covered. 

The fifth challenge involving periodogram analysis based on photometric light curves of the {\it Zwicky Transient Facility} (ZTF) is treated in \S\ref{sec:rot_methodology} below. We develop conservative and reliable extraction of rotational periodicities from Lomb-Scargle periodograms.  This results in rotational periods for 566 $[5-25]$~Myr stars with measured X-ray emission.  Although it is an increase over the A16 study of h~Per alone, it represents only a small fraction of the full multi-cluster sample of Paper~I.   

Our analysis starts with a description of the stellar samples and estimates of individual stellar masses and X-ray luminosities and other stellar properties (\S\ref{sec:other_methodology} and Appendix). Section \S\ref{sec:rot_methodology} describes methodology for obtaining reliable rotational periods from ZTF lightcurves.  The resulting rotation-mass-age relations are presented in \S\ref{sec:rotation_mass_age} and mass-stratified X-ray activity-rotation diagrams are presented in \S\ref{sec:lx_rotation}.  Section \ref{sec:spot_sizes} gives some findings relating to the sizes of star spots, a property indirectly emerging from our search for rotational periods. For our samples of e-PMS and l-PMS stars from Paper~I and the current study, \S\ref{sec:magflux} presents the connection between the inferred X-ray-rotation quantities and the data- and theory-based stellar surface magnetic flux estimates. The results are summarized in \S\ref{sec:conclusions}.

\section{Star Samples and Properties} \label{sec:other_methodology}

\subsection{Late-PMS Stellar Samples With X-ray Measurements} \label{sec:samples}

We investigate activity-rotation-age-mass relations for the {\it Gaia-Chandra} stellar members of the six $7-25$~Myr old clusters studied in Paper~I that are accessible to the Palomar~60 inch telescope used for the ZTF survey.  These include the 7~Myr old NGC 1502 (256 members from Table 4 in Paper~I), 7~Myr old NGC 2169 (104 members), 13~Myr old NGC 869 (1370 members) and NGC 884 (1294 members) commonly known as h and $\chi$ Persei, 22~Myr old NGC 1960 (406 members), and 25~Myr old NGC 2232 (117 members)\footnote{Cluster ages are obtained in Paper I by fitting Gaia EDR3 color-magnitude diagrams with isochrones from PARSEC 1.2S evolutionary models \citep{Bressan12}. Ages for larger fields of view are given in \S2.2 here.}. Out of 3547 members of these clusters, 3410 have ZTF light curves with at least 100 photometric observations that we examine in this study. 

The {\it Chandra} X-ray and Gaia-EDR3 data giving cluster memberships are discussed in Paper~I. Inferred cluster properties, including distance from the Sun, age, visual extinction, numbers of known Gaia-{\it Chandra} stellar members, and mass completeness limits for the Gaia-{\it Chandra} stellar samples are listed in Paper~I's Table~1. Field star contamination depends on cluster distance and field of view (Paper I, Figure Set 4): clusters with small fields and distances $\lesssim$1 kpc have less contamination (NGC 1502, NGC 2169, NGC 1960) while clusters with  distances $\simeq$2 kpc suffer higher  contamination (NGC 869, NGC 884, NGC 3293, NGC 4755, NGC 3766). Nearby clusters with larger fields of view have high contamination as well (IC 2395 and NGC 2232). Gaia color-magnitude diagrams are shown in their Figure Set 4. The X-ray luminosity ($L_X$) measurements and upper limits, as well as stellar properties such as Gaia-based effective temperature, mass, and bolometric luminosity, are presented in their Table 4. 

Table~\ref{tab:p_lx_ztf_and_ngc2362} here gives stellar properties and X-ray luminosities for a subset of 471 stars from Paper~I for which we obtain reliable rotational periods from ZTF light curves  (\S\ref{sec:rot_methodology}). We then add 95 stellar members of the 5~Myr old NGC~2362 open cluster, the oldest region in our earlier MYStIX (Massive Young Star-Forming Complex Study in Infrared and X-Ray) survey \citep{Feigelson13}, for which stellar rotational periods are obtained by \citet{Irwin2008}. Most X-ray measurements for NGC~2362 members are listed in Table 6 of Paper~I; additional information appears in Appendix~\ref{sec:appendix_NGC2362} below.  

The last column of Table 1 presents estimated convective turnover timescales obtained by matching stellar ages and masses to the mass-age-turnover timescale dataset from the stellar evolutionary model of \citet{Amard2019}. In cases of unstable mass regimes when the modeled $\tau_{conv}$ abruptly drops from high values of $200-300$~days to small values $<10$~days within a short mass span, $\tau_{conv}$ estimates are omitted.

A straightforward quantitative assessment of the fraction of unresolved binaries by {\it Gaia} among the 566 X-ray stars listed in Table~\ref{tab:p_lx_ztf_and_ngc2362} can be conducted by examining the quantity of {\it Gaia}-DR3 re-normalized unit weight error (RUWE). Out of the 566 stars, only 16 ($<3$\%) have a RUWE that exceeds 1.4, which is considered an indicator of poor Gaia astrometric solutions.

\begin{deluxetable*}{cccccccccccc}
\tabletypesize{\footnotesize}
\tablecaption{$5-25$~Myr PMS Stars With Rotation And X-ray Properties \label{tab:p_lx_ztf_and_ngc2362}}
\tablewidth{0pt}
\tablehead{
\colhead{Src} & \colhead{Cluster} & \colhead{R.A.} &
\colhead{Decl.} & \colhead{$\log(T_{eff})$} &
\colhead{$M$} & \colhead{$\log(L_{bol})$} & \colhead{$P_{rot}$} & \colhead{Ampl} &
\colhead{$\log(L_{X})$} & \colhead{$\log(L_{X,up})$} & \colhead{$\tau_{conv}$}\\
\colhead{} & \colhead{} & \colhead{deg} &  \colhead{deg} & \colhead{K} & \colhead{$M_{\odot}$} & \colhead{$L_{\odot}$} & \colhead{day} & \colhead{mag} & \colhead{erg s$^{-1}$} & \colhead{erg s$^{-1}$} & \colhead{day}\\
\colhead{(1)} & \colhead{(2)} & \colhead{(3)} & \colhead{(4)} & \colhead{(5)} & \colhead{(6)} & \colhead{(7)} & \colhead{(8)} & \colhead{(9)} & \colhead{(10)} & \colhead{(11)} & \colhead{(12)}
}
\startdata
1 & NGC1502 & 61.971574 & 62.278508 & 3.64 & 0.92 & -0.30 & 11.55 & 0.050 & \nodata & 28.96 & 268.3\\
2 & NGC1502 & 61.910148 & 62.277657 & 3.62 & 0.76 & -0.49 & 0.91 & 0.051 & \nodata & 28.96 & 363.1\\
3 & NGC1502 & 61.926728 & 62.368586 & 3.55 & 0.56 & -0.76 & 0.30 & 0.131 & \nodata & 28.97 & \nodata\\
4 & NGC1502 & 61.920049 & 62.261276 & 4.15 & 3.55 & 2.22 & 3.31 & 0.060 & \nodata & 28.97 & \nodata\\
5 & NGC1502 & 61.823610 & 62.298647 & 3.62 & 0.77 & -0.49 & 0.79 & 0.045 & \nodata & 28.98 & 357.9\\
6 & NGC1502 & 62.118786 & 62.365436 & 3.62 & 0.78 & -0.48 & 3.70 & 0.061 & \nodata & 28.99 & 351.9\\
7 & NGC1502 & 61.862238 & 62.242426 & 3.57 & 0.70 & -0.62 & 5.74 & 0.163 & \nodata & 29.17 & \nodata\\
8 & NGC1502 & 62.137218 & 62.374642 & 3.65 & 1.03 & -0.24 & 3.48 & 0.046 & \nodata & 29.03 & 238.0\\
9 & NGC1502 & 62.237633 & 62.260949 & 3.63 & 0.80 & -0.40 & 11.00 & 0.056 & \nodata & 29.38 & 334.1\\
10 & NGC1502 & 62.228920 & 62.387899 & 3.57 & 0.69 & -0.62 & 10.48 & 0.140 & \nodata & 29.41 & \nodata\\
\enddata 
\tablecomments{This table is available in its entirety in the machine-readable form in the online journal. Out of the 566 table records, the first 471 indicate stars with ZTF-based rotation periods, which are members of NGC 1502, NGC 2169, NGC 869, NGC 884, NGC 1960, and NGC 2232 \citep{Getman22}. The remaining 95 records present stellar members of the 5~Myr old NGC 2362 cluster, with rotation periods taken from \citet{Irwin2008}. Column descriptions: \\
Column 1: Sequential source number from 1 to 566. Column 2: Cluster name. Columns 3-4: Gaia DR3 right ascension and declination (in decimal degrees) for epoch J2000.0. Columns 5-7: Stellar effective temperature, mass, and bolometric luminosity derived from the Gaia color-magnitude diagrams. For the stars $1-471$, these quantities are from Table~4 of \citet{Getman22}. For the NGC~2362 members, these quantities are re-derived in the current paper. Column 8: Rotation period. For the stars $1-471$, these are derived here from ZTF lightcurves; values for NGC~2362 are taken from \citet{Irwin2008}. Column 9: Peak-to-peak photometric amplitude of the spot rotational modulation from folded lightcurves. Columns 10-11: $Chandra$ X-ray luminosities or upper-limits (in the $0.5-8$~keV band) from Tables~4 and 6 of \citet{Getman22}. For faint X-ray stellar members of NGC~2362, additional {\it Chandra} X-ray luminosities are derived in Appendix~\ref{sec:appendix_NGC2362}.  Column 12: Convective turnover timescale integrated over the convective envelope based on the stellar mass, age and rotation using  models of \citet{Amard2019}.}
\end{deluxetable*}

\subsection{Additional Late-PMS Stars Without X-ray Measurements}
\label{sec:other_rotation_masses}

For the purpose of comparison with modern rotation evolution models (\S \ref{sec:rotation_mass_age}) and analyses of stellar spot areas (\S \ref{sec:spot_sizes}), we extend our X-ray-rotation data sample (\S \ref{sec:samples}, Table~\ref{tab:p_lx_ztf_and_ngc2362}) by adding other stars in the $5-25$~Myr age range (mostly without $Chandra$ data) with previously reported rotation periods. These additional stars include members of the NGC~2362 and NGC~869 (including many members outside $Chandra$ fields), Upper Sco (US), Upper Centaurus-Lupus (UCL) and Lower Centaurus-Crux (LCC) subregions of the Sco-Cen Association, and the Pleiades. Their membership and rotation period information are taken from \citet{Irwin2008}, \citet{Moraux2013}, \citet{Rebull2018}, \citet{Rebull2022}, and \citet{Rebull2016}. For US, UCL, and LCC, we only consider the `gold' members with a single rotation period from Rebull et al. (2018, 2022). 

In order to place all these stars on the same age-mass scale as the stars from our X-ray-rotation sample, we apply PARSEC 1.2S evolutionary models \citep{Bressan12,Chen14} to Gaia-DR3 data \citep{GaiaDR32022} using the same procedures as with ZTF clusters (Paper~I) and NGC~2362 MYStIX stars (Appendix~\ref{sec:appendix_NGC2362}). Specifically, we match these stars to the Gaia-DR3 catalog using a radius of $1\arcsec$, estimate median cluster distances from the Sun using only stars with $RUWE<1.4$ and $\sigma_{\omega} < 0.1$~mag, truncate the samples to include only stars with the Gaia color uncertainty of $\sigma_{BP} < 0.1$~mag, create Gaia-DR3 color-magnitude diagrams, fit the color-magnitude distributions with the reddened PARSEC 1.2S models to derive cluster extinction/age and stellar masses/temperatures/luminosities. 

The Gaia-inferred cluster distance, extinction, and age are as follows: NGC~2362 ($d=1301$~pc, $A_V=0.3$~mag, $t=5$~Myr as in Appendix \ref{sec:appendix_NGC2362}), US ($d=142$~pc, $A_V=0.7$~mag, $t=9$~Myr), NGC~869 ($d=2451$~pc, $A_V=1.7$~mag, $t=13$~Myr as in Paper~I), UCL ($d=142$~pc, $A_V=0.3$~mag, $t=15$~Myr), LCC ($d=114$~pc, $A_V=0.1$~mag, $t=15$~Myr), and the Pleiades ($d=$136~pc, $A_V=0.2$~mag, $t=100$~Myr). Appendix~\ref{sec:appendix_m_p} tabulates these additional stars and shows Gaia color-magnitude diagrams. 

\subsection{X-ray Luminosities} \label{sec:xray_luminosities}

For many thousands of stellar members of very young (generally $t<5$~Myr) star forming regions, such as the MYStIX regions mentioned in \S \ref{sec:samples}, Table~6 of Paper~I  lists their {\it Chandra}-based intrinsic X-ray luminosities in the $0.5-8$~keV band. These quantities were derived using the non-parametric method XPHOT \citep{Getman10} based on a concept similar to the use of color–magnitude diagrams in optical and infrared astronomy. The XPHOT method requires the presence of both soft ($<2$~keV) and hard ($>2$~keV) X-ray photons.

But for the six older and less absorbed clusters with ZTF stellar rotation periods derived here, hard X-ray photons are often lacking.  Paper~I estimated stellar intrinsic X-ray luminosities by stacking all individual stellar {\it Chandra}  spectra (typically of low S/N) into one high-S/N spectrum, obtaining intrinsic X-ray luminosity ($L_X$) for the merged spectrum.  This merged $L_X$ is divided by the sum of apparent X-ray fluxes of  the involved stars to estimate a conversion factor from intrinsic to apparent fluxes which is then applied to estimate intrinsic X-ray luminosities of individual X-ray detected cluster members. In addition, for the Gaia cluster members with no X-ray detections,  Paper~I estimates upper limits on $L_X$ based on spatial positions in the {\it Chandra} ACIS-I detector. These X-ray luminosities and upper limits for the members of the six ZTF clusters are listed in Table~\ref{tab:p_lx_ztf_and_ngc2362}.  Treatment of NGC~2362 members is further discussed in Appendix \ref{sec:appendix_NGC2362}. 

With the inclusion of the Gaia cluster members with X-ray non-detections, the full {\it Chandra-Gaia} stellar samples of the six ZTF clusters listed in Table~4 of \citet{Getman22} are complete down to masses of $0.4-0.7$~$M_{\odot}$ for NGC 1502, NGC 2169, NGC 1960, and NGC 2232, and down to $1$~$M_{\odot}$ for the more distant NGC 869 and NGC 884 clusters. 

The X-ray-rotation stellar samples given in our Table~\ref{tab:p_lx_ztf_and_ngc2362} can not be considered complete for any mass stratum due to the absence of reliable ZTF rotational periods for most $Chandra-Gaia$ stars.  We thus stress that any X-ray-rotation correlations or slopes observed/measured in the following sections may be biased with respect to any intrinsic relations.

\section{Photometric Rotational Periods from ZTF Light Curves}
\label{sec:rot_methodology}

The Zwicky Transient Facility (ZTF) is an automated wide-field survey of $\sim 30,000$ square degrees of the northern sky obtained with the 48-inch Oschin Schmidt telescope at Palomar Observatory.    The ZTF survey is described by  \citet{Masci19} and \citet{Bellm19} with access provided by the NASA/IPAC Infrared Science Archive.  Our analysis is based on Data Release 10 (DR10).

After matching the 3547 {\it Chandra}-Gaia l-PMS cluster members discussed in \S\ref{sec:samples} to the ZTF point source catalog using a matching radius of 2\arcsec, we select 3410 ZTF light curves with at least 100 photometric observations in the $r$ or $g$ bands, after flagging for bad photometry. Unreliable data points are further removed with two rounds of `sigma clipping’ using a $4 \times IQR$ (InterQuartile Range) criterion.  For some analysis steps, the irregular cadence is converted to a regular cadence with 1.0 day bin width. Note that we did not attempt to use the more sophisticated custom ZTF data reduction method by \citet{Lu2022}, which involves considering archival ZTF light curves for multiple stars located near the target star to remove the systematics that are common to all these stars.

Before proceeding to results, we discuss the difficulties of measuring rotation periods from irregularly spaced light curves obtained from a single telescope site (\S\ref{sec:period_reliability}).  We then present our methodology in some detail, as it differs from other treatments in the field (\S\ref{sec:ztf_periodicity_determination}).

\subsection{Obtaining Reliable Photometric Rotation Periods in PMS Stars} 
\label{sec:period_reliability}

Fully convective e-PMS stars often exhibit $\sim 100$~mmag rotational modulation of large cool star spots, allowing photometric determination of their rotational periods \citep[e.g.][]{Bouvier93, Herbst00, Affer13}.  The amplitude of periodic variability can range over a factor of 10.  The effect declines during the l-PMS phase; in  Upper Sco, many stars have spot amplitudes around $30-80$~mmag \citep{Rebull2020AJ}.  Reported rotation periods mostly fall into the range $\sim 1-10$~day with a few extending to $15-20$ days during both e-PMS and l-PMS phases.  Photometric measurements of rotation throughout the early phases of stellar evolution, together with observable bulk stellar properties (mass, radius, age), give insights into surface magnetic activity and internal dynamo processes. 

However, photometric rotation periods of PMS stars are not easily measured. First, the datasets have significant deficiencies.  NASA's 4-year Kepler observations did not include the Galactic Plane.  The K2 mission covered parts of the Galactic Plane with $\sim 3$ month light curves but without associated sensitive X-ray measurements.  TESS Full-Frame Image light curves  are subject to crowding and nebulosity effects and have not yet been analyzed for PMS stars.  Periods will be limited to $\lesssim 10$ days due for single-sector observations.   Ground-based surveys such as HATSouth and ZTF have hundreds of observations spanning several years, but their photometric accuracy is subject to atmospheric and instrumental limitations.  

Second, star spot instability can confuse periodic rotational effects for multi-year light curves.  e-PMS star spots may remain large and stationary on the photosphere for years but may also suddenly change \citep{Vrba88, Grankin09}.  Spot durations have been estimated for a few l-PMS stars with durations in the range $\sim 100-400$~days \citep{Maldonado2022}. But in the Sun, spots often fade in weeks and are replaced by spots at different longitudes.  The resulting changes in phase, even with a constant period, broadens and weakens the peak in a periodogram.  

Third, many stellar light curves from  sensitive space-based photometry reveal autocorrelated stellar variations from magnetic activity, stellar pulsations or super-granulation motions \citep{Caceres19b}.  Autocorrelated variations from atmospheric conditions must be removed from ground-based photometry using algorithms like SysRem and the Trend Filtering Algorithm \citep{Tamuz05, Kovacs05} prior to a search for stellar rotational periodicity.

Fourth, the statistical methodology for obtaining rotational periods from irregular light curves has considerable limitations. The most common tool is the Lomb-Scargle periodogram (LSP), the Fourier power spectrum modified for irregularly spaced cadences \citep{Scargle82}.  Even in the best of circumstances, the evaluation of significance of LSP peaks $-$ commonly called False Alarm Probabilities $-$ is difficult to evaluate \citep[e.g.][]{Koen90, Baluev08, Suveges14, Delisle20}. 

\begin{figure*}[ht!]
\includegraphics[width=\textwidth]{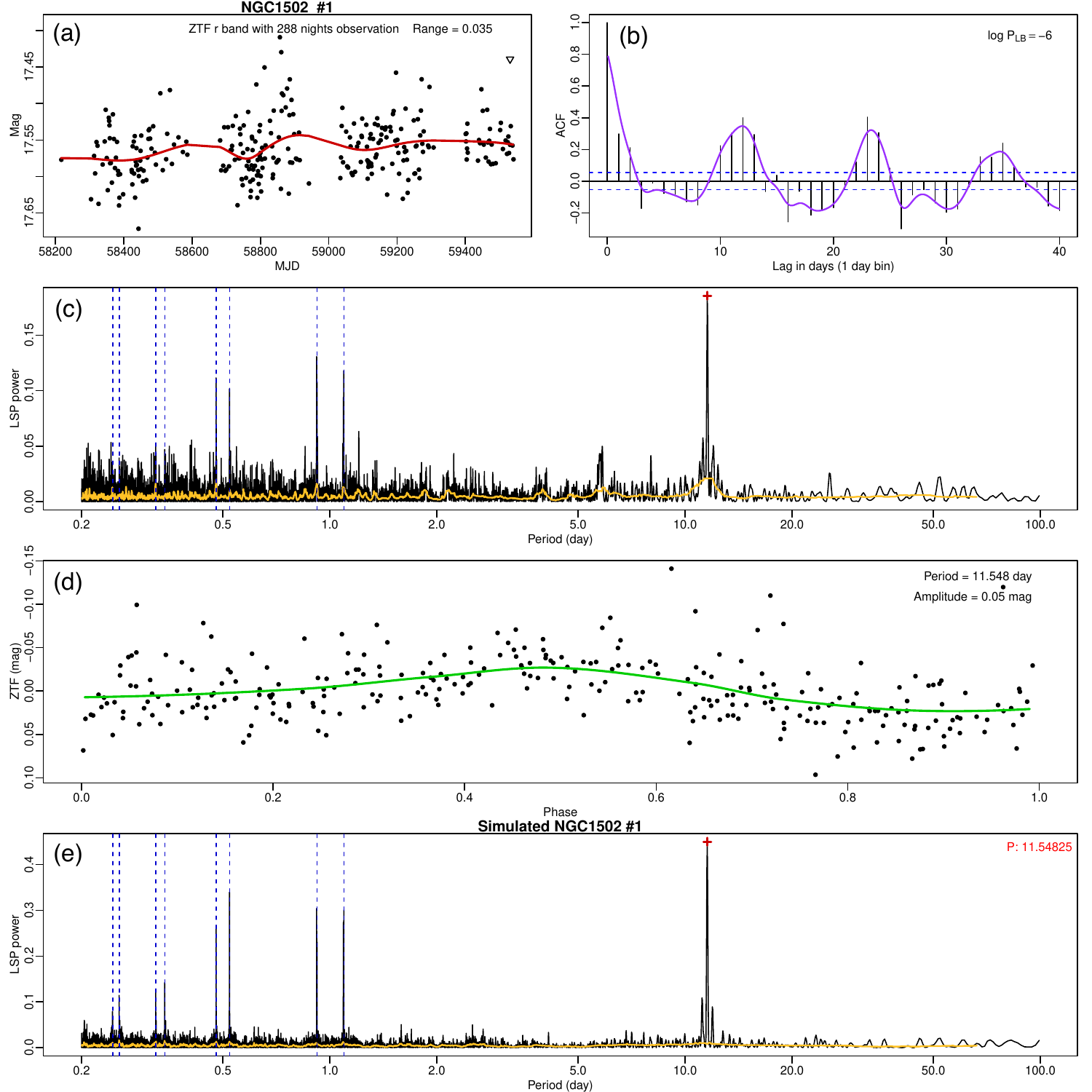}
\caption{Results from the rotational period search for 3410 ZTF light curves. This is the first page from a figure set for 471 stars with reliable ZTF rotation periods available in the electronic version of the paper. Panel (a) ZTF lightcurve. The legend gives the number of ZTF observation nights and magnitude range of the lightcurve.  Panel (b) Autocorrelation function after collecting the light curve into 1~day bins. The log of the Ljung-Box probability is shown; cases with $\log(P_{LB}) <  -2$ require autocorrelation while cases with $\log(P_{LB}) > -2$ are consistent with white noise. Panel (c) Lomb-Scargle periodogram for the unbinned ZTF data. The red plus marks the chosen best period, and dashed lines show expected diurnal aliases \citep{VanderPlas2018}. Panel (d) Folded lightcurve for the best LSP period. The green curve gives a robust local polynomial fit to the data using R's {\it loess} algorithm (see text for details). The legend gives the period and amplitude of a smooth local regression fit.  Panel (e) Lomb-Scargle periodogram for a simulation of the best rotation period and amplitude, plus Gaussian noise, using the original ZTF observation times.} \label{fig:ztf_analyses_results}
\end{figure*}

But the situation is considerably worse for ground-based surveys from a single observatory where the star is visible for only several hours each day and several months each year due to solar motion.  When the light curves are folded with periods around 1.0 days or its aliases, sinusoidal fits have unstable amplitudes leading to false and fluctuating power in the periodogram \citep{Dawson10, VanderPlas2018}.  Figure \ref{fig:ztf_analyses_results} shows an example of the Lomb-Scargle periodogram for a spotted l-PMS star with rotation period of 11.5~days observed with ZTF. But strong peaks are also seen at 1.00, 0.50 and 0.25 days from the diurnal solar cycle. Peaks may also appear around 28 days from the lunar cycle.  These structures in the periodogram can be present even if no true stellar periodicity is present.  Most of these peaks are spurious and attempts to apply straightforward False Alarm Probabilities to these irregular cadences will often fail.

As an alternative to the LSP, the nonparametric autocorrelation function (ACF) is sometimes used for detecting rotational periods.  \citet{McQuillan14} argue that the first peak of the ACF gives a reliable estimate of the rotational period.  But there is no statistical guidance on how to construct the ACF: different first peaks will appear with different bin widths or smoothers used to display the ACF. First peaks may also arise from noise or aperiodic autoregressive variations rather than meaningful rotational signals.  

In the present study, we use both the LSP and ACF methods.  We  test possible periodicities with simulated light curves reproducing the peculiar observational cadence of each light curve.  The existence of a possible rotational signal is then evaluated by conservative subjective judgment: 
\begin{enumerate}
\item An LSP peak stronger than spurious solar and lunar peaks must be present.
\item A strong periodic signal must be present in the ACF if the period is longer than $\sim 3$~days.
\item The LSP and ACF periods must agree, understanding that the LSP period is more accurate than the ACF period. 
\end{enumerate}  
\vspace*{-0.1in}

\begin{figure*}[ht!]
\includegraphics[width=\textwidth]{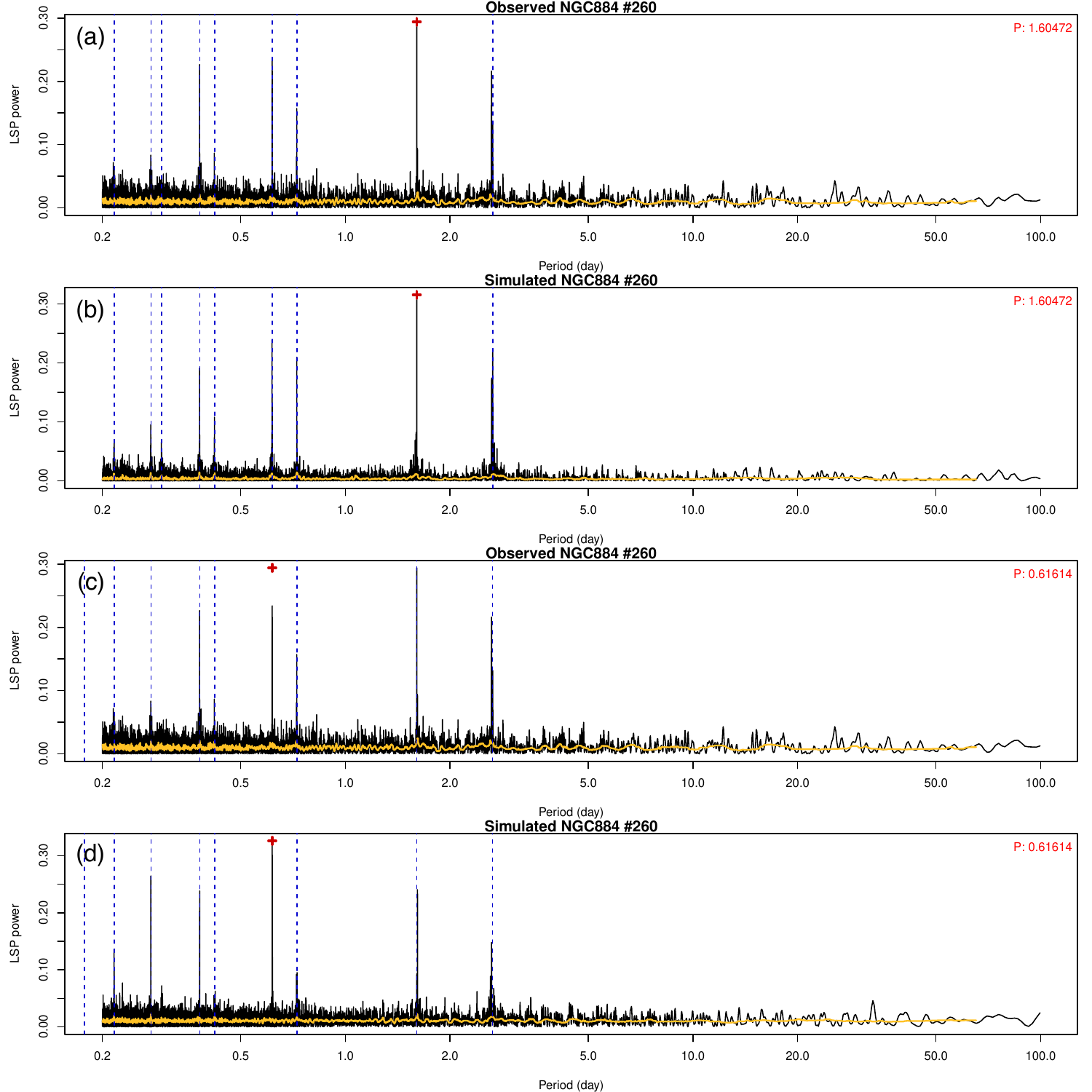}
\caption{Lomb-Scargle periodograms for source \#260 in NGC~884. Panels a and c show the same LSP for the observed ZTF lightcurve but assuming two different ``true'' periods, 1.60 and 0.61 days (marked by the red pluses). Corresponding expected failure modes are marked by the blue dashed lines. Panels b and d show the results of simulations for the trials of the two assumed periods.  The simulation trial with the 1.60 day period more closely resembles the real periodogram.}  \label{fig:observations_vs_simulations}
\end{figure*}

\begin{figure*}[t]
\centering
\includegraphics[width=0.65\textwidth]{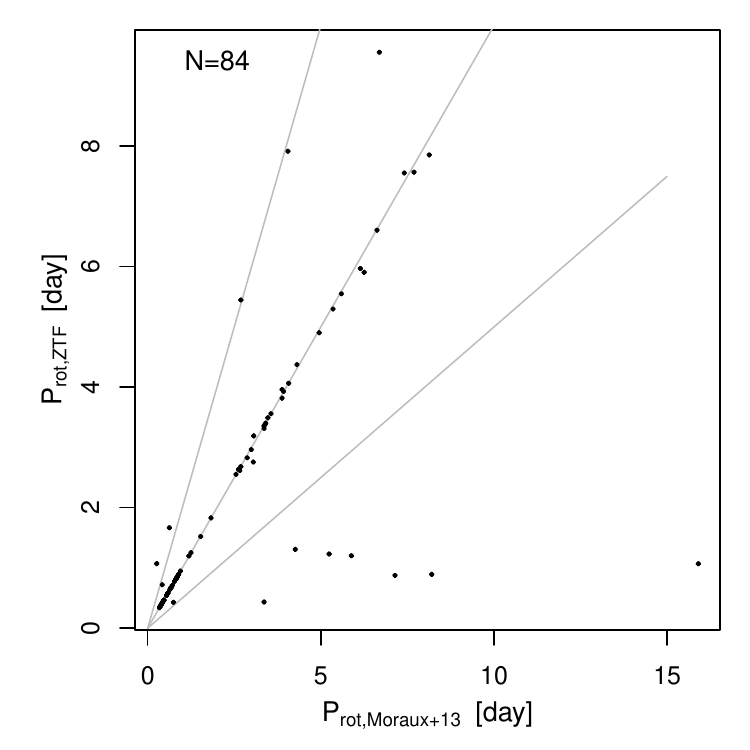}
\caption{Comparison between our ZTF periods and previously published rotation periods by \citet{Moraux2013} for 84 stars in NGC~869 (= h Per). The grey lines mark the unity, 1/2$\times$ and 2$\times$ unity lines. }  \label{fig:p_comparison_ngc869}
\end{figure*}

\subsection{Periodogram Analysis} \label{sec:ztf_periodicity_determination}

The Lomb-Scargle periodogram is calculated using CRAN package $lomb$ in the R public domain statistical software environment \citep{RCoreTeam21, Ruf99} based on the algorithm of \citet{VanderPlas15}. Our LSP examines periods ranging from 0.2 to 100 days.   No attempt is made to estimate statistical significance of peaks given the difficulties outlined in \S\ref{sec:period_reliability} and illustrated in Figures~\ref{fig:ztf_analyses_results} and \ref{fig:observations_vs_simulations}. 

The autocorrelation function is calculated using R’s $acf$ function \citep{VenablesRipley2002} applied to the 1-day binned light curve. ACFs constructed with smaller bins were unsuccessful due to the proliferation of missing pairs of measurements at short lags. As with the Discrete Correlation Function of \citet{Edelson88}, missing contributions to the correlation function due to cadence gaps are ignored.  Confidence intervals corresponding to 95\% probabilities assuming uncorrelated Gaussian noise and no missing values are shown as dashed lines in Figure~\ref{fig:ztf_analyses_results}b.  While they underestimate the true confidence intervals given the erratic number of missing values in each lag, they are useful to give a rough indication of ACF features with low statistical significance.  More usefully, the probability that no autocorrelation is present is calculated with the Ljung-Box omnibus hypothesis test for autocorrelation at any lag time  \citep{LjungBox1978}.  The Ljung-Box test generalizes the classical Durbin-Watson test for serial (lag=1 time unit) autocorrelation.  We calculate $P_{LB}$ using R's $Box.test$ function with the Ljung-Box test option.

Based on visual inspection of resulting LSPs and ACFs, we select an initial sub-sample of 657 light curves from the original sample of 3410 ZTF stars with very low values of $\log(P_{LB})$ and high signal-to-noise LSP periodograms.  These light curves show one or more periodic signals. We find that the ACFs are insensitive to short rotation periods but are often consistent with one of the dominant LSP peaks in cases of long rotation periods.  

Due to the interaction between observation's window function and spectral power, the resulting LSPs exhibit spurious harmonics and aliases that researchers may mistakenly interpret as true rotation periods \citep{Dawson10, VanderPlas2018}. For the vast majority of our stars, the ZTF-based LSPs show the presence of strong peaks within both low-period ($P \lesssim 1.5$~day) and high-period ($P \gtrsim 1.5$~day) ranges. Based on simulations, we often find that one of the peaks, often but not always with the highest LSP power, is the true period while the others are spurious.   

In our simulations, using R's $sample$ and $sin$ functions, we first scramble the ZTF photometric measurements to create randomized datasets with erased periodicity or other autocorrelated structure.  Artificial sinusoidal signals are then injected with LSP trial periods with amplitudes obtained from the {\it loess} fits to the folded ligthcurves\footnote{The well-established `loess' (LOcally Estimated Scatterplot Smoothing) regression model, also known as the Savitsky-Golay filter, fits quadratic functions in local windows that pass sequentially with increasing rotational phase \citep{Cleveland88}. The model includes robust downweighting of outliers using M estimation with Tukey's biweight function.  The method is implemented in R function {\it loess}.}.  For each choice, we mark expected harmonics for $m=1$ and $n= \pm (1,2,3,4)$ according to equations (45)-(47) from \citet{VanderPlas2018}. For all 657 stars with high S/N periodograms, we ran two trials of simulations. In the first (second) trial, we assume that the strongest spectral peak at short periods $P<1.5$~day (long periods $P>1.5$~day) is the true period. The simulation trial where the alias structure closely matches that of the observed ZTF LSP is considered to be the true period. 

Figure~\ref{fig:ztf_analyses_results} shows an example with a clear long period seen in both the ACF and LSP. It is a sample page of the electronic Figure Set that presents the ZTF lightcurve, ACF, LSP, folded lightcurve and simulation trial for the best period solution (marked with red cross).   Figure~\ref{fig:observations_vs_simulations} shows a more challenging case. All trial periods are too short to be detected with the ACF, and two trial periods give similar predicted aliases.  However, the alias strengths of the simulated LSPs for $P=1.60$~day more closely match the observed alias structure than the trial at $P=0.61$~day.  

These simulations allowed us to select clear best periods for 471 out of 657 ZTF stars with strong LSP peaks. Only these 471 ZTF stars are tabulated and used in our X-ray activity-rotation analyses below; true periodic behaviors are probably present in other stars but we cannot identify the individual periods in a reliable fashion.  The final 471 ZTF-based periods, and associated amplitudes of the variation in smoothed folded lightcurves, are listed in Table~\ref{tab:p_lx_ztf_and_ngc2362}.  

The last 95 lines of the table list the NGC 2362 stars in common between the {\it Chandra} X-ray \citep{Broos13, Getman22} and CTIO Blanco/Mosaic-II based rotation period \citep{Irwin2008} samples. Here the rotation periods and amplitudes of folded light curves are from \citet{Irwin2008}; stellar masses, effective temperatures, bolometric and X-ray luminosities are from \S\ref{sec:appendix_NGC2362}.

Among the 471 ZTF stars with reliable rotational periods, 154 are members of the central part of the NGC~869 (= h~Per) stellar cluster covered in the {\it Chandra} survey \citep{Getman22}. Previously, based on their multi-telescope optical monitoring of a larger area of NGC~869, \citet{Moraux2013} published rotation periods for 586 candidate cluster members. There are 84 stars that are common between our current ZTF analysis and the Moraux et al. study. Figure \ref{fig:p_comparison_ngc869} shows that the ZTF periods are generally consistent with the periods (or half-periods) of \citet{Moraux2013}, with the exception of seven stars, for which our simulations prefer shorter-period solutions. Similar discrepancies between stellar rotational periods derived from different datasets analyzed with different methods are shown in \citet{Briegal22}. 

Finally we note that some periodic photometric signals that we interpret to be rotationally modulated spotted single stars may actually be eclipsing binary systems with a F- or G-type primary and an M-type secondary \citep{Triaud13}.  We suspect this source of contamination is not common, as our target stars are restricted to X-ray and kinematic cluster members (Paper I) rather than drawn from a large population of field stars.  

\section{Skumanich-Like Relations for late-PMS stars} \label{sec:resuts}

\subsection{ Rotation Dependence on Mass and Age} \label{sec:rotation_mass_age}

\begin{figure*}[ht!]
\epsscale{1.15}
\plotone{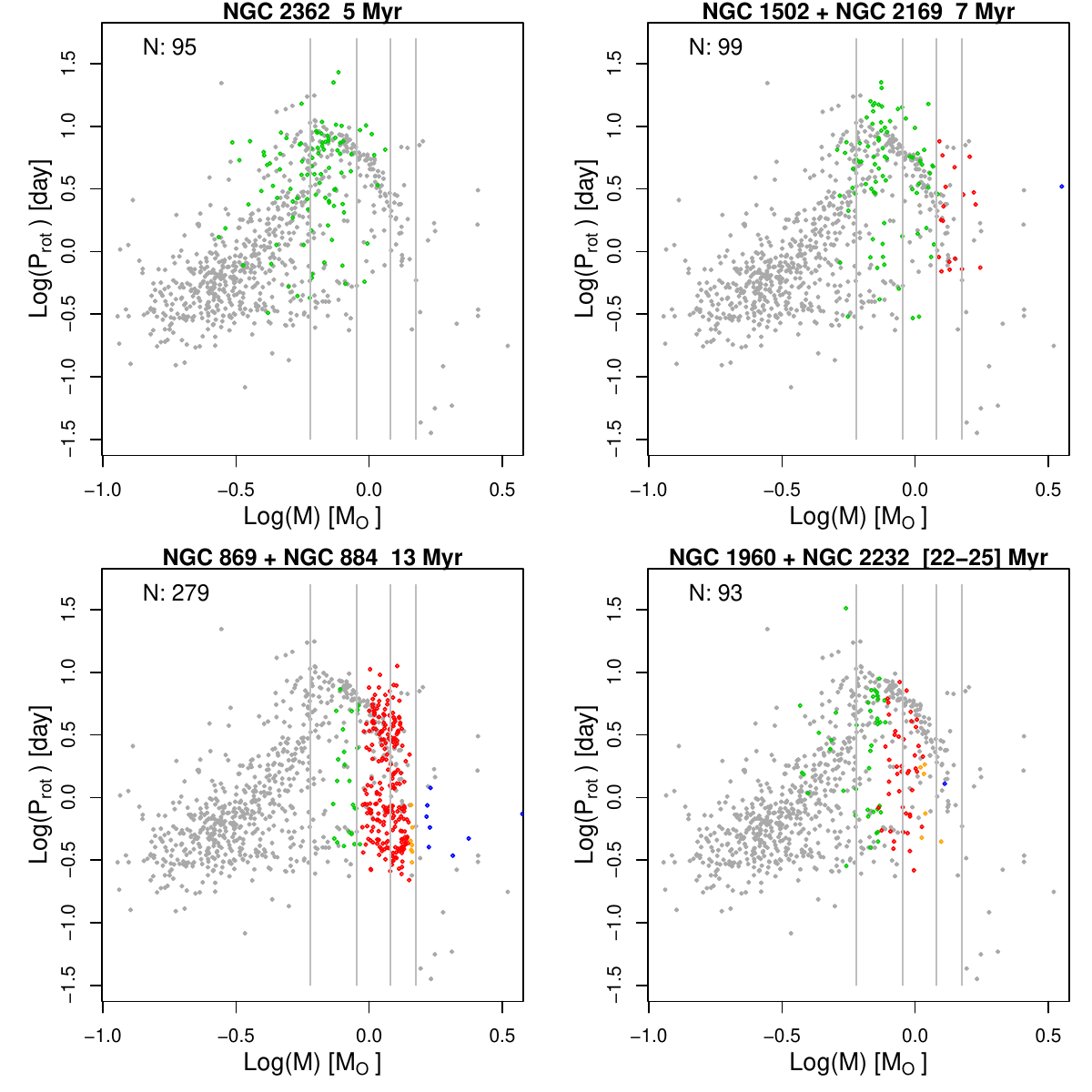}
\caption{Rotation-mass diagrams for seven l-PMS clusters (colored points) compared to members of the Pleiades cluster \citep[grey points,][see Appendix B]{Rebull2016}. The X-ray-rotation clusters are arranged into 4 age strata, and individual stars are color-coded by their evolutionary phase on the Hertzprung-Russell diagram:  Hayashi track (green; $ph_{HRD}<2$), early Henyey track (red; $2 \leq ph_{HRD} <3$), late Henyey track (orange; $3 \leq ph_{HRD} <4$), and ZAMS-MS  (blue; $4 \leq ph_{HRD} <6$). The three mass strata discussed in the text are marked by the grey lines. Figure legends present numbers of plotted X-ray-rotation stars.}  \label{fig:logp_logm}
\end{figure*}

For the X-ray-rotation sample in Table~\ref{tab:p_lx_ztf_and_ngc2362}, Figure~\ref{fig:logp_logm} displays the distribution of rotation periods as a function of stellar mass in three mass strata $-$ $(0.6-0.9)$, $(0.9-1.2)$, and $(1.2-1.5)$~M$_{\odot}$ $-$ and four cluster age strata: 5 Myr, 7 Myr, 13 Myr and 22-25 Myr.  A handful of stars lie beyond $M < 0.6$~M$_{\odot}$ and $M > 1.5$~M$_{\odot}$.   The l-PMS cluster distributions are compared to the well-studied Pleiades ZAMS-MS stars. At a given cluster age, the mass strata distinguish between stars of different evolutionary phases, such as Hayashi versus Henyey tracks at 7~Myr, 13~Myr, and $>20$~Myr. These evolutionary phases, called $ph_{HRD}$ here, are taken from the PARSEC 1.2S mass tracks by \citet{Chen14}. Appendix \S\ref{sec:appendix_hrd_phase} provides further details and tabulation of such phases for all l-PMS stars from Paper~I. 

\begin{figure*}[hb]
\centering
\includegraphics[width=0.95\textwidth]{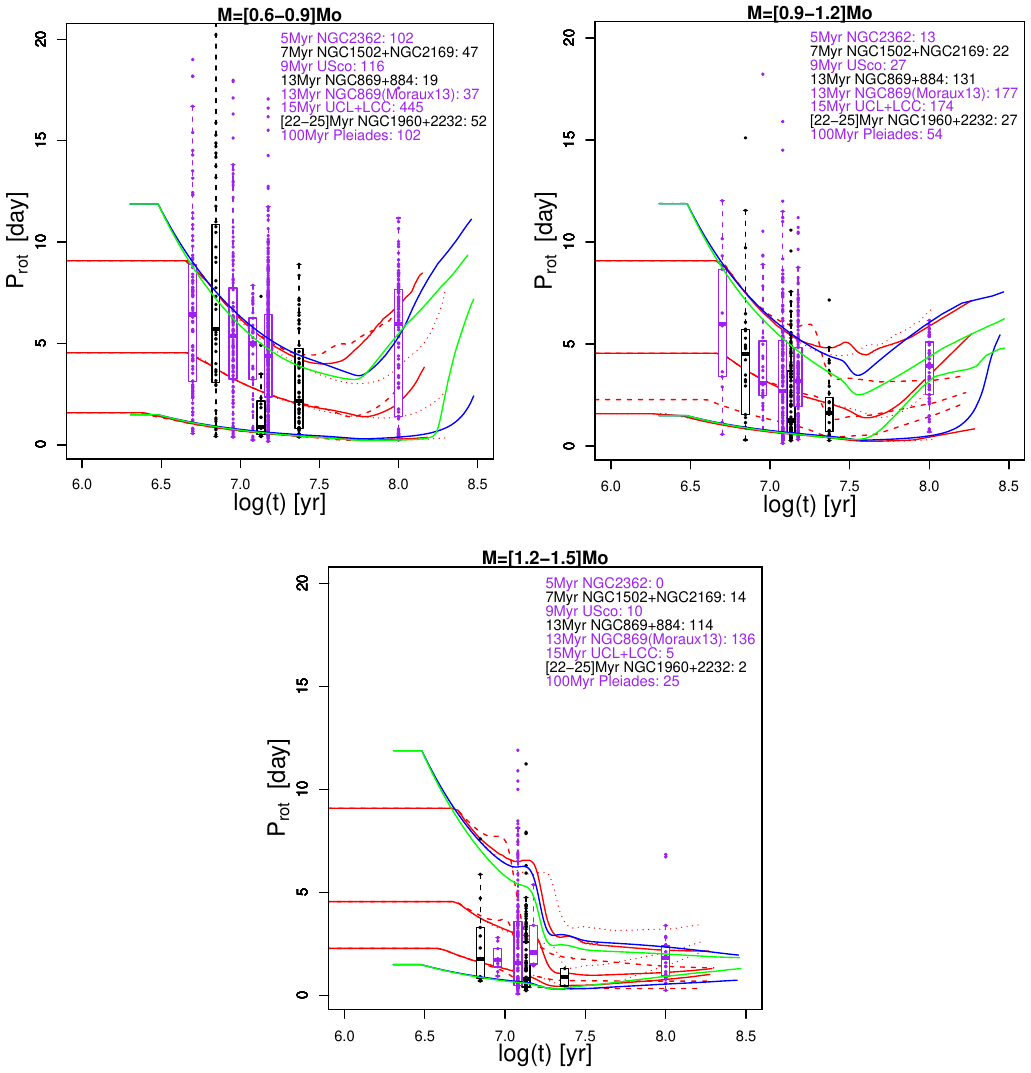}
\caption{Observed and calculated evolution of stellar rotation. Black (Table 1) and magenta (previously published) dots show rotational periods with box-and-whisker diagrams for l-PMS clusters.  Colored curves give the rotation evolutionary tracks from \citet{Amard2019} and \citet{Gossage2021} (see text for details).  Figure legends indicate the cluster names and number of plotted stars.} \label{fig:comparison_with_models}
\end{figure*}

An initial result from Figure~\ref{fig:logp_logm} is that while some are more rapidly rotating than Pleiades stars, others have quickly spun down to the $P_{rot}-M$ locus seen during ZAMS stage.  This is seen on both the Hayashi and Henyey tracks.  In these slower rotating stars, the changes in internal structure during the l-PMS phase do not cause strong spin-up at the surface. This result was found by \citet{Moraux2013} who state: ``at least a fraction of slow rotators are prevented from spinning up'' during the l-PMS phase.  We caution that our methodology restricts period determination from ZTF lightcurves to only reliable cases (\S\ref{sec:rot_methodology}), it is quite likely that some fast rotators are missing from our sample.  For example, rapid l-PMS rotators may produce multi-modal surface field morphologies with smaller star spots rather than one large hemispheric star spot needed for reliable photometric detection. 

We note that the slower rotators ($2 < P_{rot} < 10$ days) in our l-PMS sample show greater scatter than the narrow locus of slow rotators in the Pleiades in the $P_{rot}-M$ diagrams.  This is not due to observational uncertainties, as $P_{rot}$ values are measured with high accuracy from our periodograms.  This suggests that the angular momenta of an ensemble of l-PMS stars have not yet settled down to the well-defined rotational patterns seen on the ZAMS. 

The $P_{rot}$  distribution can be compared to modern stellar evolutionary models with self-consistent treatment of rotation internal angular momentum transport and rotational mixing processes \citep{Amard2019, Gossage2021}.  Gossage et al. present models with two  magnetic effects: traditional magnetic braking of \citet{Matt2015} and consideration of the morphology of stellar surface magnetic fields by \citet{Garraffo2018}.

Figure~\ref{fig:comparison_with_models} compares the temporal evolution of stellar rotation predicted by these models with our rotational data. Individual stars from our X-ray-rotation ZTF sample (Table~\ref{tab:p_lx_ztf_and_ngc2362}) and their associated box-and-whisker diagrams are shown in black while stars with previously published rotation periods (Table~\ref{tab:gaia_masses_previous_stars})  are shown in magenta. The colored curves give the rotation evolutionary tracks.  Models from \citet{Amard2019} are in red for three different initial rotation rates (slow, median, fast) and three different masses per figure panel ($0.6,0.8,0.9$~M$_{\odot}$, $0.9,1,1.2$~M$_{\odot}$, and $1.2,1.3,1.5$~M$_{\odot}$). Models from \citet{Gossage2021} are shown for magnetic braking prescriptions of \citet[][in blue]{Matt2015} and \citet[][in green]{Garraffo2018} for a single mass per figure panel ($0.8,1,1.3$~M$_{\odot}$) and for two initial rotation rates (fast and slow).  During the e-PMS phase at ages $\lesssim 4$~Myr,  the models assume stellar rotation rates are constant due to disk coupling. Figure legends indicate numbers of plotted stars.

Figure~\ref{fig:comparison_with_models} shows broad agreement between observations and theory. The median and quartile rotation rates increases following disk decoupling during $7-25$~Myr are due to contraction along the Hayashi and Henyey tracks. The spin-up is particularly notable in our NGC~1960 and NGC~2232 stars at ages $22-25$~Myr.  Contraction slows during $30-100$~Myr and rotation slows on the main sequence due to centrifugal momentum loss by a magnetic wind. The predicted U-shaped pattern of rotation periods is seen in all mass strata with the fastest rotations seen in NGC 1960 and NGC 2232.   

Some discrepancy, however, can be noted.  Particularly for the lower $0.6-0.9$~M$_{\odot}$ mass stratum, a quarter or more of the observed rotations are slower than the rotation model tracks: the models are missing a very slow rotator population. This could be be attributed to disk locking timescales substantially longer than $\sim 4$~Myr in many lower-mass stars \citep{Ribas2015, Richert18}.  It thus does not represent a fundamental disagreement between data and models. 

The bifurcation in $P_{rot}$ distribution in the NGC~869 cluster reported by \citet{Moraux2013} is confirmed using our independent rotation data.  It is also present in the rich NGC~884 cluster; it appears as a deficit of stars with $P_{rot} \simeq 1$ day (Figures \ref{fig:logp_logm} and \ref{fig:comparison_with_models}).
 
For the $0.9-1.2$~M$_{\odot}$ mass stratum we compare the distributions of rotation periods for the rich samples of similar aged UCL/LCC and NGC~869 clusters taken from space-based data of \citet{Rebull2022} and ground-based data of \citet{Moraux2013}, respectively. Using the Anderson-Darling 2-sample test, the period distribution of \citet{Moraux2013} is significantly different from UCL ($p<0.001$) and is probably different from LCC ($p=0.02$).  UCL and LCC have indistinguishable period distributions.  The NGC~869 and UCL/LCC differences are mainly due to the difference in spread rather than in median.  The nonparametric Ansari-Bradley test, where the alternative hypothesis assumes same medians but different spreads,  give $p<0.001$ that they have the same spread. Compared to the space-based UCL/LCC data, the NGC~869 data exhibit an excess of rapidly rotating stars.

Moraux et al. suggest that the excess may be due to a population of tidally locked close binary systems identified by their photometric locus in a color-magnitude diagram.  The bifurcation in rotational periods thus may not reflect a stage in the rotational evolution of single stars.  Alternatively, the existence of very rapid rotators in l-PMS clusters with $P_{rot}$ as short as 8 hours may be related to the truncation of protoplanetary disks in wide binary stars \citep{Meibom07}.  

However, while \citet{Rebull2022} derive space-based rotation periods for the vast majority of the UCL/LCC cluster members, the NGC~869 data of \citet{Moraux2013} and our own NGC~869+884 ZTF data are highly incomplete in the determination of the rotation periods. We thus can not evaluate the true fraction of rapid rotators in these clusters as well as their nature.
  
\subsection{X-ray Dependence on Mass, Rotation and Age} \label{sec:lx_rotation}

We use the stellar sample from Table~\ref{tab:p_lx_ztf_and_ngc2362} to study relationships between X-ray luminosity and stellar rotation, particularly to examine whether supersaturated, saturated and unsaturated regimes are present in $5-25$~Myr l-PMS stars as they are in ZAMS stars.

\begin{figure*}
\centering
\includegraphics[width=0.93\textwidth]{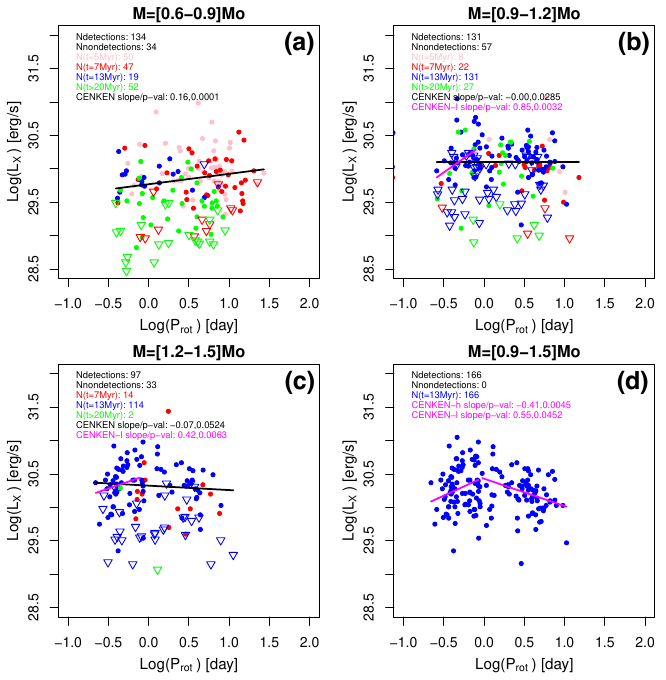}
\caption{X-ray luminosity as a function of rotation period for the l-PMS stars in Table~\ref{tab:p_lx_ztf_and_ngc2362}. Panels (a,b,c) show $L_X-P_{rot}$ distributions for different mass strata.  X-ray detections (circles) and X-ray non-detections (triangles) are color-coded according to stellar age: $5$~Myr (pink), $7$~Myr (red), $13$~Myr (blue), and $>20$~Myr (green). Black and magenta lines are calculated using the Akritas-Thiel-Sen statistical procedure; see text for detail.  Panel (d) shows $L_X-P_{rot}$ distribution for 13~Myr old stars that were detected in X-rays in the $0.9-1.5$~M$_{\odot}$ range.} \label{fig:lx_p}
\end{figure*}

Figure~\ref{fig:lx_p} shows the l-PMS mass-stratified $L_X-P_{rot}$ distributions. As the X-ray data consist of a mixture of detections and upper limits, we use methods of survival analysis to evaluate the presence of (anti)correlation \citep{Feigelson2012}.  We calculate Helsel's generalized Kendall's tau $p$-value for null hypothesis of no correlation, and  the Akritas-Theil-Sen linear regression slope \citep{Helsel2012,Akritas1995}.  These are implemented with functions {\it cenken} and {\it akr} in CRAN package {\it NADA} within the R statistical software environment \citep{Lee2020}.   We looked for correlations in mass-stratified $L_X - P_{rot}$ diagrams for  individual clusters as well as cluster ensembles, for all rotation periods as well rotations constrained to rapid ($P_{rot} \la 1$~day) and slow ($P_{rot} \ga 1$~day) ranges with different boundaries. 

In the $0.6-0.9$~M$_{\odot}$ stratum (dominated by Hayashi-track stars), stellar samples for individual clusters do not show any $L_X-P_{rot}$ correlations, but the cluster ensemble sample exhibits a weak positive correlation with slope $L_X \propto P_{rot}^{0.2}$ (black line in Figure~\ref{fig:lx_p}a). The generalized Kendall's $\tau$ p-value is $p \simeq 0.0001$. This positive correlation in l-PMS stars is opposite to the well-known anticorrelation between $L_X$ and $P_{rot}$ for unsaturated ZAMS stars.  But the l-PMS effect may be coincidental.  $L_X$ and $P_{rot}$ decay with age for different reasons: X-ray luminosity falls as the distributed dynamo weakens \citep{Getman22} while rotation period declines as contracting stars spin up following disk decoupling (Figure~\ref{fig:comparison_with_models}).

In the $0.9-1.2$ and $1.2-1.5$~M$_{\odot}$ mass strata, our X-ray-rotation samples are dominated by the Henyey-track stellar members of $13$~Myr old, rich NGC~$869$ and NGC~$884$ clusters (Figure~\ref{fig:lx_p}b and c). Other individual clusters are represented by sparse samples with no $L_X-P_{rot}$ correlations. For the entire rotation range, using either the entire sample of all open clusters (as shown in figure panels b and c) or just a sub-sample of the dominant $13$~Myr old clusters, $L_X$ appears independent of $P_{rot}$ (Kendall's $\tau$ p-value $p > 0.03$). But one can construct a sub-sample of rapidly rotating l-PMS stars where $L_X$ increases with $P_{rot}$ (magenta lines, $p \simeq 0.005$). This is consistent with the supersaturation result of A16 for $1.0-1.4$~M$_\odot$ stars in NGC~869.  

In Figure~\ref{fig:lx_p} panel d where only X-ray detections are considered (to mimic the X-ray sensitivity level of the shallower {\it Chandra} observation from A16), we find relations $L_X \propto P_{rot}^{0.6}$ for $P_{rot} < 1$~day ($p \simeq 0.05$, which is not significant) and $L_X \propto P_{rot}^{-0.4}$ ($p \simeq 0.005$, which is significant) for $P_{rot} > 1$~day.  The latter is similar to the non-saturation result of A16. 

A16 interpret these results as the onset of the supersaturated and unsaturated tachoclinal dynamo regimes that dominate the ZAMS phase.  

But we should also emphasize strong selection effects in our sample of l-PMS stars in Figure~\ref{fig:lx_p}.  While the underlying Gaia-$Chandra$ samples of cluster members are large and reasonably complete for solar-mass stars, only $\simeq 13$\% have reliable rotational periods from our ZTF periodogram analysis. Only stars with large hemisphere-size star spots have reliable rotational periods. The ZTF-Gaia-{\it Chandra} sub-sample lacks weaker X-ray stars, so the distribution of X-ray luminosities differs considerably from the full Gaia-$Chandra$ sample for NGC~869$+$884 (Paper~I).  It is further possible that our rotational periods for NGC~869 and NGC~884 stars are missing long-period stars that were found by \citet{Moraux2013} (Figure~\ref{fig:comparison_with_models}) and is thereby unable to reveal the unsaturated anticorrelation between $L_X$ and $P_{rot}$.  We thus conclude that our findings in Figure~\ref{fig:lx_p} are consistent with, but not a convincing independent demonstration of results reported by A16. 

We note that the relation $L_X \propto P_{rot}^{-0.4}$ for $P_{rot} > 1$~day shown in Figure~\ref{fig:lx_p}d is much shallower than the slope of $\simeq -2$ measured for main sequence stars \citep{Pizzolato2003, Wright2011}. Furthermore, the longest rotation periods of $\sim 10$~days for our l-PMS stars correspond to Rossby number of $Ro = P/\tau_{conv} \sim 0.1$ (Table~\ref{tab:p_lx_ztf_and_ngc2362}) whereas the unsaturated  $L_X$-rotation regime in older stars starts at $Ro \ga 0.13$.  It may be more realistic to view the $L_X \propto P_{rot}^{-0.4}$ effect found here and by A16 as a mild (factor of 2-3) decrease of X-ray activity within a saturated dynamo regime. 

\clearpage\newpage

\section{Late Pre-Main Sequence Star Spot Sizes}
\label{sec:spot_sizes}

A byproduct of photometric searches for rotational periods is a measurement of the fractional surface area occupied by large star spots.  If the stellar radius is known from Gaia multiband photometry and parallax measurements, then the star's spot area can also be calculated.  This quantity is not evaluated from the periodogram where the LSP peak power is difficult to interpret, but from the light curve folded modulo the period selected by the process described in \S\ref{sec:rot_methodology}.  We discuss here star spot sizes as a measure of surface activity in addition to X-ray luminosity.  A forthcoming study will examine the utility of star spot sizes to estimate ages for PMS and ZAMS stars. 

Star spot fractional coverages are obtained from $loess$ local regression fit to folded ZTF light curves, as shown by the green curve in panel d of Figure~\ref{fig:ztf_analyses_results}. The spot coverage fraction is the peak-to-peak amplitude of this fitted function in magnitudes.  The measured amplitudes for the ZTF lightcurves are listed in column 9 of Table~\ref{tab:p_lx_ztf_and_ngc2362}.  Previously reported amplitudes by \citet{Irwin2008} and \citet{Moraux2013} for NGC~2362 and NGC~869 stars with Gaia-based masses are listed in Table~\ref{tab:gaia_masses_previous_stars}. 

Spot areas are calculated assuming that the observed periodic photometric is caused by rotational modulation of a single dominant spot (or a conglomerate of nearby smaller spots).  Spot sizes are estimated following \citet{Herbst2021}.  Spot area $A_{spot}$ is proportional to the observed drop in brightness relative to the average stellar flux ($\Delta F/F_{av}$), projected stellar hemisphere area ($A_{star} = \pi R_{star}^2$), and stellar ($T_{eff}$) and spot ($T_{spot}$) temperatures:  
\begin{equation} \label{eq:spot_area}
A_{spot} = A_{star} \Delta F/F_{av} (1-(T_{spot}/T_{eff})^4)^{-1}
\end{equation} 
where the stellar and spot temperatures are related as $T_{spot} = -3.58 \times 10^{-5} T_{eff}^2 + 1.0188 T_{eff} - 239.3$. $\Delta F/F_{av}$ is obtained from the peak-to-peak magnitude amplitude ($Ampl$) as $\Delta F/F_{av} = 10^{0.2 Ampl} - 10^{-0.2 Ampl} \approx Ampl$.  See \citet{Kounkel2022} for a related calculation.

\begin{figure}
\centering
\includegraphics[width=0.5\textwidth]{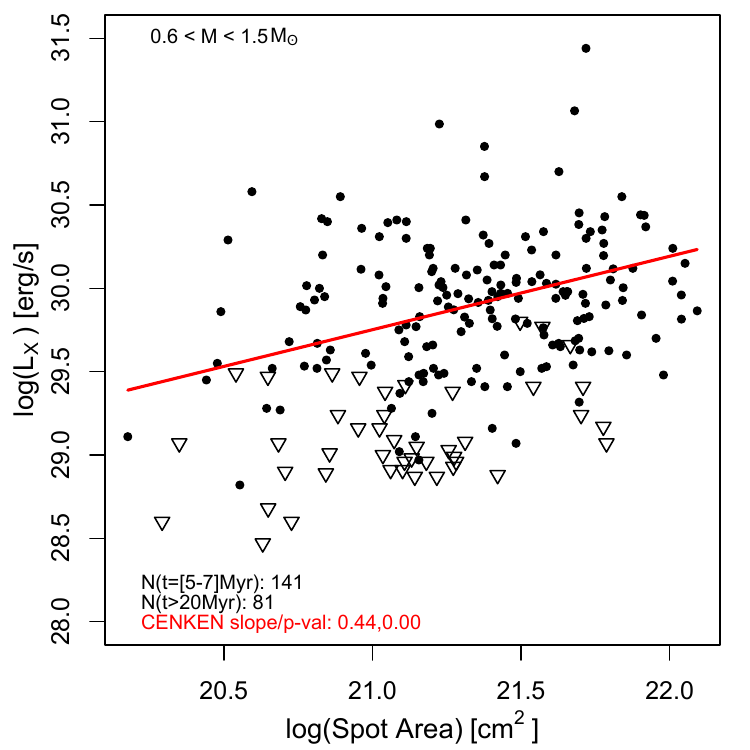}
\caption{X-ray luminosity as a function of surface spot area for l-PMS stars.  The sample in this plot is restricted to stars with ages $t=[5-7]$~Myr and $t>20$~Myr (omitting NGC 869 and NGC 884 stars at 13~Myr stars) and masses $0.6-1.5$~M$_{\odot}$. X-ray detections and upper limits are marked by the solid circles and triangles, respectively. The red line shows the Akritas-Thiel-Sen slope.} \label{fig:lx_vs_spotarea_individual_stars}
\end{figure}

Figure~\ref{fig:lx_vs_spotarea_individual_stars} shows that X-ray luminosity and star spot area are strongly correlated as $L_X \propto A_{spot}^{0.4}$ using the Akritas-Thiel-Sen procedure.  NGC~869 and NGC~884 stars, dominating our X-ray-rotation sample, are omitted from this plot since their limited $L_X$ and $A_{spot}$ dynamic ranges and low sensitivity towards $<0.9$~M$_{\odot}$ stars weakens the correlation.  If these clusters are included, the dependence is still statistically significant but is weaker, $L_X \propto A_{spot}^{0.1}$.  The power law index ranges from 0.3 to 0.6 when other mass ranges or subsamples are considered.

We believe this is the first demonstration that X-ray emission and spot sizes are correlated measures of surface magnetic activity in a stellar population. However, the derived slope should not be interpreted astrophysically because our X-ray-rotation stellar samples from Table~\ref{tab:p_lx_ztf_and_ngc2362} are badly incomplete due to the absence of reliable ZTF rotational periods for most {\it Chandra}-Gaia stars from Paper~I. The spot areas are only measurable in ZTF periodograms when the star spot area is both large and distributed asymmetrically across the surface so that one hemisphere is substantially darkened compared to the other hemisphere.

\begin{figure*}[ht!]
\includegraphics[width=0.95\textwidth]{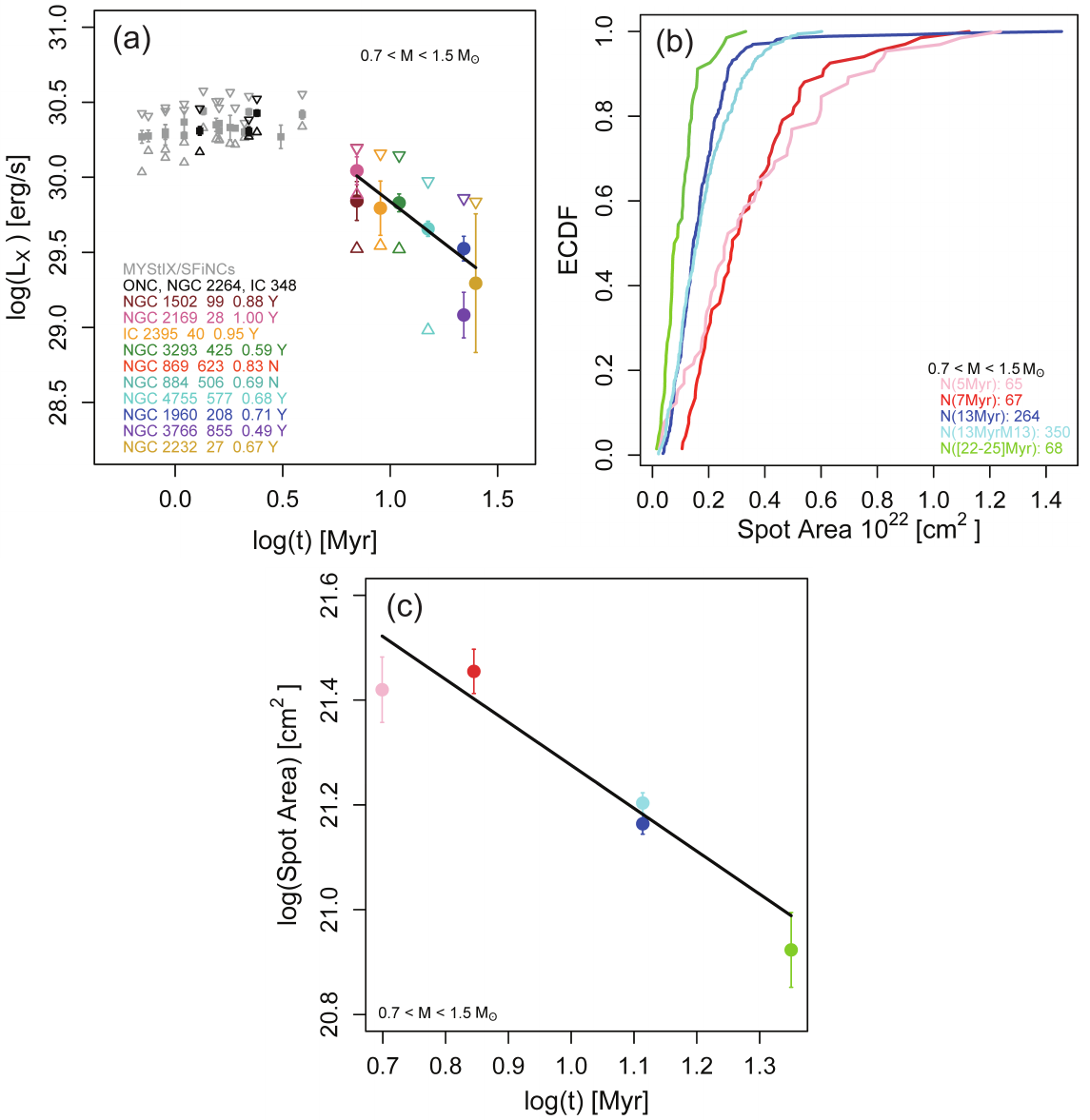}
\caption{Temporal evolution of X-ray luminosity and spot area in late PMS stars. (a) Evolution of X-ray luminosity through the e-PMS and l-PMS phases, adapted from Paper~I \citep[][see their Figure~8 for details]{Getman22}. (b) Empirical cumulative distribution function of spot area for four age strata: 5~Myr (pink), 7~Myr (red), 13~Myr (dark blue), and $22-25$~Myr (green).  The 13~Myr sample from \citet{Moraux2013} is shown for comparison (light blue). (c) For the distributions in panel (b), median values with bootstrapped 68\% confidence intervals with weighted linear fit. The samples for panels (b,c) are drawn from Tables~\ref{tab:p_lx_ztf_and_ngc2362} and \ref{tab:gaia_masses_previous_stars} of the current paper.} 
\label{fig:lx_aspot_temp_evol}
\end{figure*}

More insight might emerge from independent examination of how $L_X$ and $A_{spot}$ evolve during the l-PMS phase in Figure~\ref{fig:lx_aspot_temp_evol}.  The $L_X-t$ plot in panel a is taken from Paper~I where the decay slope during the l-PMS phase was derived employing all known {\it Chandra}-Gaia stellar members of open clusters that have mass completeness limits $\la 0.7$~M$_{\odot}$. The evolution of spot area in panels b and c are based on all stars with known peak-to-peak amplitudes given in Tables \ref{tab:p_lx_ztf_and_ngc2362} and \ref{tab:gaia_masses_previous_stars}.

\begin{deluxetable*}{cccccccc}
\tabletypesize{\footnotesize}
\tablecaption{Temporal Evolution of X-ray Luminosity and Spot Area \label{tab:lx_aspot_slopes}}
\tablewidth{0pt}
\tablehead{
\colhead{Mass} & \colhead{$N(L_X)$} & \colhead{$b$} &
\colhead{$p$-val(b)} & \colhead{$N(A_{spot})$} & \colhead{$a$} &
\colhead{$p$-val(a)} &  \colhead{$b/a$}\\
\colhead{$M_{\odot}$} & \colhead{} & \colhead{} &  \colhead{} & \colhead{} & \colhead{} & \colhead{} & \colhead{ }\\
\colhead{(1)} & \colhead{(2)} & \colhead{(3)} & \colhead{(4)} & \colhead{(5)} & \colhead{(6)} & \colhead{(7)} & \colhead{(8)}
}
\startdata
$0.7-0.9$ & 556 & $-0.52 \pm 0.22$ & 0.07 & 178 & $-0.70 \pm 0.33$ & 0.12 & $0.74 \pm 0.47$\\
$0.9-1.2$ & 845 & $-0.61 \pm 0.16$ & 0.01 & 370 & $-0.88 \pm 0.10$ & 0.00 & $0.69 \pm 0.20$\\
$0.7-1.2$ & 1769 & $-0.83 \pm 0.30$ & 0.03 & 548 & $-0.77 \pm 0.17$ & 0.02 & $1.08 \pm 0.46$\\
$0.9-1.5$ & 1335 & $-1.16 \pm 0.15$ & 0.00 & 636 & $-0.99 \pm 0.17$ & 0.01 & $1.17 \pm 0.25$\\
$0.7-1.5$ & 2259 & $-1.10 \pm 0.24$ & 0.00 & 814 & $-0.85 \pm 0.16$ & 0.01 & $1.29 \pm 0.37$\\
\enddata 
\tablecomments{Column 1: Mass range. Column 2: Number of PMS stars evolved in the analysis of the temporal evolution of $L_X$ using data and methods from Paper~I. Columns 3-4:  Results from the linear regression fits for the relation $L_X \sim t^b$. The results include the slope ($b$) with 68\% error, and p-value for the hypothesis of zero slope. Column 5: Number of PMS stars evolved in the analysis of the temporal evolution of spot area ($A_{spot}$). Columns 6-7: Results from the linear regression fits for the relation $A_{spot} \sim t^a$. The results include the slope ($a$) with 68\% error, and p-value for the hypothesis of zero slope. Column 8: Inferred slopes and 68\% errors for the relation $L_X \sim A_{spot}^{b/a}$.}
\end{deluxetable*}

Figure~\ref{fig:lx_aspot_temp_evol} demonstrates that spot area, like X-ray luminosity, drops several-fold during the l-PMS phase.  The plots show that $A_{spot} \propto t^a$ with $a \simeq -0.9$ and $L_X \propto t^b$ with $b \simeq -1.1$ for the $0.7-1.5$~M$_{\odot}$ mass range.  Evolutionary trends for other mass ranges are listed in Table~\ref{tab:lx_aspot_slopes}.  These results are consistent among different rotation datasets: for ages $5-7$~Myr, the spot areas are similar for our ZTF (red point in Figure~\ref{fig:lx_aspot_temp_evol}c) and CTIO-based data/analyses of \citet[][pink point]{Irwin2008}; and for age $13$~Myr, our ZTF spot areas (blue point) agree with \citet[][cyan point]{Moraux2013}.  

Figure~\ref{fig:lx_aspot_temp_evol}c dramatically shows that the decline of the median stellar spot area with age follows a power law relation.

\begin{figure}
\centering
\includegraphics[width=0.5\textwidth]{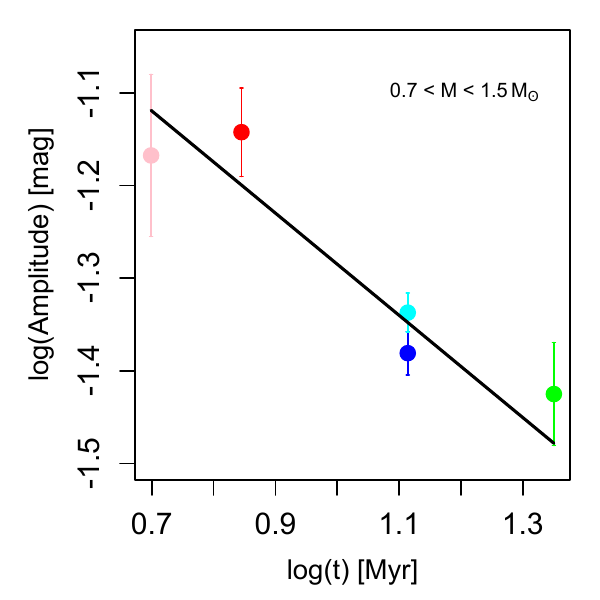}
\caption{Same as in Figure \ref{fig:lx_aspot_temp_evol}c but for the amplitude of the photometric variations.} \label{fig:amplitude}
\end{figure}

Some studies of the evolution of star spots \citep[e.g.][]{Rebull2018} use the amplitude of the photometric variations rather than the spot area.  In our sample of $5-25$~Myr stars when stellar radii can vary substantially, we find that the terms $A_{star}$ and $Ampl$ in equation \ref{eq:spot_area} also decay with time, but slower than $A_{spot}$. For instance, Figure~\ref{fig:amplitude} shows that for the $(0.7-1.5)$~M$_{\odot}$ stellar sample from Figure~\ref{fig:lx_aspot_temp_evol}b,c the  amplitude decreases with time as $Ampl = t^{-0.53 \pm 0.15}$. The stellar hemisphere area decreases approximately as $A_{star} \sim t^{-0.3}$, while $A_{spot} \propto t^{-0.9}$ (Table~\ref{tab:lx_aspot_slopes}).  This temporal evolution of $A_{spot}$ supports the magnetic activity paradigm (see below). Spot area is thus an effective measure of star spot evolution in this l-PMS regime. 

Hemispheric spot area fraction ($f_{spot} = A_{spot}/A_{star}$) decays approximately as $f_{spot} \sim t^{-0.5}$. This is consistent with the Kepler and TESS study of six clusters with ages ranging from $10$~Myr to 4~Gyr \citep{Morris2020}, for which the authors report $f_{spot} \sim t^c$ with $c = -0.37 \pm 0.16$.


Table~\ref{tab:lx_aspot_slopes} shows that both $A_{spot}$ and $L_X$ decrease approximately as $t^{-1.0}$ for solar mass stars with a somewhat shallower drop around $t^{-0.7}$ for sub-solar mass stars. 

We thus find that the time-averaged X-ray luminosity and surface spot area of l-PMS stars are nearly proportional to each other with $b/a \simeq 1$. A plausible explanation is that local spot magnetic field strengths are roughly constant but the areas of spots vary and evolve among l-PMS stars.  This is motivated by Zeeman broadening measurements in solar-type stars from PMS to main sequence.  They show a constant spot magnetic field at $B \sim 3$~kG but magnetic filling factor ($f$), and thus disk-averaged magnetic field strength $<B>=Bf$, decreasing with age \citep{Kochukhov2020, Sokal2020}. For our l-PMS stars, assuming that the spot magnetic field strength is also constant, the $L_X \propto A_{spot}^1$ relationship (Figure~\ref{fig:lx_aspot_temp_evol} and Table~\ref{tab:lx_aspot_slopes}) implies that the time-averaged X-ray luminosity would scale linearly with disk-averaged stellar magnetic flux $L_X \propto \Phi \sim B \times f_{spot} \times R^{2} \sim B \times A_{spot}$. The implied $L_X \propto \Phi$ scaling appears to be universal in the Sun and stars (\S \ref{sec:magflux}).  This scaling would imply that $L_X \propto A_{spot}^1$ would be seen in l-PMS stars, or any sample of normal stars. ~\\

\begin{figure*}[ht!]
\epsscale{1.15}
\plotone{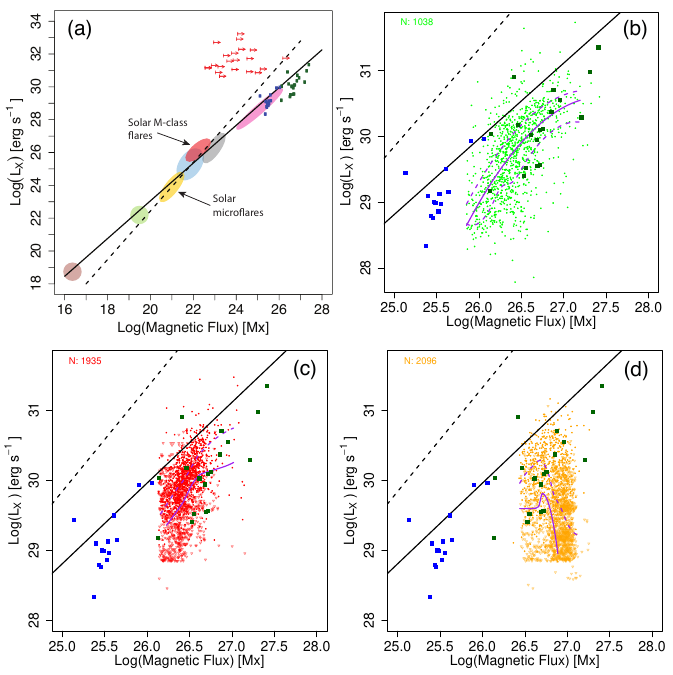}
\caption{The Pevtsov relation linking X-ray luminosity to magnetic flux. Panel (a) is reproduced from \citet{Getman2021b}. The solid and dashed black lines are the relations $L_X \sim \Phi^m$, where $m=1.15$ \citep{Pevtsov2003} and $m=1.48$ \citep{Kirichenko2017}, respectively. Magnetic structures include the quiet Sun (brown), solar X-ray-bright points (light green), solar active regions (cyan), solar disk averages (gray), and old G, K, and M dwarfs (magenta). The solar microflares (golden oval) are from \citet{Kirichenko2017} and solar M-class flares (red oval) are from \citet{Su2007}. Individual solar-type dwarfs (blue squares) are from \citet{Kochukhov2020}. Individual Orion and Taurus e-PMS stars (dark green squares) are from \citet{Sokal2020}. Panels (b), (c) and (d) expand the high-luminosity regime with the \citet{Pevtsov2003} and \citet{Kirichenko2017} regression lines shown. Panel (b) shows e-PMS stellar members of Orion Nebula, IC~348, and NGC~2264 from Table~6 of Paper~I (light green) with kinetic magnetic flux estimated according to equation  \ref{eq:magnetic_strength}. All these stars are on Hayashi tracks ($ph_{HRD} < 2$). The purple curves are spline fits to the 25\%, 50\%, and 75\% quartiles of the $L_X-\Phi$ distribution (see text for details). Panels (c) and (d) show l-PMS stars from Paper~I with stars on early Henyey tracks in red ($2.0 \leq ph_{HRD} <2.7$) and stars on later Henyey and ZAMS-MS tracks in orange ($2.7 \leq ph_{HRD} <6.0$). The purple curves show smoothed 50\% and 75\% Kaplan-Meier quartiles of the $L_X-\Phi$ distributions (see text for details).} 
\label{fig:fig_discussion}
\end{figure*}

\section{X-ray Emission and Surface Magnetic Flux} \label{sec:magflux}

Two decades ago, \citet{Pevtsov2003} found a remarkable power law relationship that links solar and stellar magnetic events; a version is shown in Figure~\ref{fig:fig_discussion}a.  Here the characteristic level of X-ray luminosity $L_X$ scales with the magnetic flux integrated over the responsible surface region of area $A$, $\Phi = B \times A$, according to $L_X \propto \Phi^m$ with $m=1.0-1.5$ \citep{Pevtsov2003,Kirichenko2017}\footnote{
The individual early PMS (dark green squares) and older solar-type stars (blue squares) shown in Figure~\ref{fig:fig_discussion}a have magnetic fluxes based on Zeeman broadening measurements. Zeeman Doppler imaging measurements produce a slope $m \ga 1.5$ while Zeeman broadening follow the standard slope \citep{Vidotto2014, Reiners2022}. It is unclear whether the slope difference is arises from methodology  or is intrinsic to the stars \citep{Kochukhov2020,Kochukhov2021}. 
}.  It points to a universal mechanism of coronal plasma heating from a supply of magnetic flux to the stellar surface, regardless of the nature of underlying magnetic dynamos. The relationship is present from the smallest solar structures to powerful solar and stellar flares.   

There are indications that the relationship changes for highly  magnetically active conditions. For instance, the X-ray super-flares that frequently occur in e-PMS stars may require a steeper slope with $m \simeq 3$ and unusually strong local surface fields of up to 20~kG \citep{Getman2021b, Zhuleku2021}.

For rapidly rotating stars with convection-driven dynamos such as e-PMS or mid-M dwarfs, the available kinetic energy of convective flows may set the strength of the surface magnetic field \citep{Christensen2009, Reiners2009, Reiners2010, Reiners2022}. These researchers provide a simple analytic formula for the average strength of surface magnetic field, where magnetic pressure ($\propto B^2$) depends on the stellar density $\rho^{1/3}$ (where $\rho \sim M/R^3$) and stellar bolometric flux $q^{2/3}$ (where $q \sim L/R^2$). Normalized to the empirical measurements of magnetic field strength for various stars and planets in units of $G$, the expression is 
\begin{equation} \label{eq:magnetic_strength}
B_{kin} = 4800 (M L^2/R^7)^{1/6}
\end{equation}     
where  $M$, $L$, and $R$ are stellar mass, luminosity, and radius in solar units.  

We applied this formula to fully and partially convective l-PMS Gaia-{\it Chandra} stars listed in Tables~6 and 4 of Paper~I. For comparison to e-PMS stars, we choose a sub-sample of 1038 lightly absorbed ($A_V<5$~mag) members of the nearby IC~348, Orion Nebula, and NGC~2264 star forming regions with masses $0.2-2$~M$_{\odot}$. These are the nearest, rich regions for which we have most sensitive {\it Chandra} X-ray data. These e-PMS stars are shown in Figure~\ref{fig:fig_discussion}b (light green) with quantile regression curves (purple)\footnote{
These curves are generated with the Constrained B-Splines method that combines spline and quantile regression fits.  The 50\% (median) curve is shown with a solid purple curve and 75\% and 25\% quartiles are marked by dashed curves. The procedure uses quadratic splines with knot selection based on the Akaike Information Criterion.  The method is implemented in the R CRAN {\it cobs} package \citep{Ng2007,Ng2020}. 
}. The fits show that e-PMS {\it Chandra} X-ray luminosities and kinetic magnetic surface fluxes ($B_{kin} \times 4 \pi R^2$ from equation \ref{eq:magnetic_strength}) roughly follow the universal relationship of \citet{Pevtsov2003} and \citet{Kirichenko2017} though with X-ray luminosities reduced by factors of $2-3$. A linear slope of $m=1.3$ resulting from an independent linear regression fit to these e-PMS data (not shown) supports this finding.

We similarly apply equation (\ref{eq:magnetic_strength}) to the l-PMS members of open clusters in Table~4 of Paper~I.  Here the  Gaia-{\it Chandra} sub-samples used are complete down to $\simeq 0.7$~M$_{\odot}$, not $\simeq 0.2$~M$_\odot$ as in the e-PMS case. Like their younger e-PMS siblings, numerous fully-convective l-PMS stars lying on Hayashi tracks ($ph_{HRD} < 2.0$) follow the universal law (figure is not shown). So do the l-PMS stars that lie on the early segments of Henyey tracks ($2.0 \leq ph_{HRD} < 2.7$).  These are shown as red dots in Figure~\ref{fig:fig_discussion}c with spline fits\footnote{
The l-PMS samples in  Figure~\ref{fig:fig_discussion}c and d include a substantial number of X-ray upper limits (shown as open trianges) as well as detections (shown as filled circles). To treat the limits in a statistically consistent fashion, we calculate the nonparametric maximum likelihood Kaplan-Meier estimator \citep[KM;][]{KaplanMeier58} for the $L_X$ distributions in sliding bins of $\Phi$. Local quadratic fits are then calculated to the 50\% and 75\% quartiles of the KM estimators to generate smooth curves.  Due to high fractions of X-ray nondetections in many $\Phi$ bins, the unreliable KM 25\% quartiles are omitted. Software calculations are made in R with the {\it survival} \citep{Therneau20} and {\it loess} \citep{Cleveland1992} functions. 
}. 

However, more massive l-PMS stars on the more evolved Henyey track and on the ZAMS-MS show a different behavior.  Stars on later segments of the Henyey tracks have evolutionary phases of $2.7 < ph_{HRD} < 4.0$ in Appendix \ref{sec:appendix_hrd_phase}, and ZAMS-MS stars have $4 < ph_{HRD} < 6$.  These stars, shown as orange dots in the $L_X-\Phi$ diagram of Figure~\ref{fig:fig_discussion}d with spline fits, exhibit an $anti-correlation$ between $L_X$ and $\Phi$, much different from the universal $correlation$ found by \citet{Pevtsov2003,Kirichenko2017} and seen in our younger l-PMS and e-PMS stars in Figure~\ref{fig:fig_discussion}b and c. 

As the universal $L_X-\Phi$ relationship is present for both e-PMS and main sequence stars, we do not believe it is absent for l-PMS stars. Our finding in \S\ref{sec:spot_sizes} that $L_X \propto A_{spot}^{1}$ (Table~\ref{tab:lx_aspot_slopes}) further supports a correlation between X-ray emission and surface magnetic flux levels.  Rather, the anti-correlation seen in  Figure~\ref{fig:fig_discussion}d is arguably due to a failure of equation (\ref{eq:magnetic_strength}) in estimating magnetic field strengths based on a convective dynamo.  This provides indirect evidence that the dynamos during the later stages of l-PMS and ZAMS evolution are switching from $\alpha^2$ to $\alpha\Omega$ processes.   

This change in $L_X-\Phi$ behavior occurs approximately two-thirds of the duration between the base of the Henyey track and the time of maximum bolometric luminosity on the Henyey track before descent to the ZAMS.  Recall that these results are based on PARSEC 1.2S evolutionary models.

This result agrees with different arguments in Paper~I that pointed to the  $\alpha \Omega$ dynamo as dominant in these more massive and older l-PMS stars.  Paper~I found that the X-ray luminosity of  $1-3.5$~M$_{\odot}$ stars (most of which lie on late Henyey or ZAMS tracks) decays as $L_X \propto t^{-1.8}$. This decay rate is consistent with that of main sequence stars, whose magnetic activity is driven by the tachoclinal $\alpha \Omega$ dynamo \citep{Maggio1987,Gudel1997,Feigelson2004}.

\section{Conclusions} \label{sec:conclusions}

\subsection{Observational Accomplishments}

From an observational viewpoint, this study together with Paper~I has greatly increased our knowledge of stellar activity and rotation for solar-type stars in the elusive late-pre-main sequence phase of stellar evolution.  We generate samples of several thousand stars within open clusters with ages $5-25$~Myr.  Combining Gaia and $Chandra$ observations allows cluster membership to be established with high reliability. Paper~I presents a catalog of $\sim 8000$ l-PMS stars ($\sim 6000$ and $\sim 2000$ lying in the cluster centers and outskirts, respectively), most of which have X-ray detections.   

In this study, we examined photometric light curves for 3410 of these l-PMS using the northern hemisphere multi-epoch survey of the Zwicky Transient Facility.  We establish a rigorous procedure for identifying reliable photometric periodicities due to rotationally modulated star spots using light curves obtained at a single telescope site. This analysis gives rotation periods for 471 Gaia-{\it Chandra}-ZTF stars ranging from around 0.5 to 8 days (\S\ref{sec:rot_methodology}). Several dozen stellar members of NGC~2362 with {\it Chandra}-CTIO measurements were added to this X-ray rotation sample (Appendix \ref{sec:appendix_NGC2362} and Table~\ref{tab:p_lx_ztf_and_ngc2362}). For some analyses,
we also consider previously published rotation periods for additional $>3500$ stars, mostly without {\it Chandra} data (Appendix \ref{sec:appendix_m_p}).

The resulting mass-dependent evolution of rotation through the l-PMS phase agrees well with astrophysical models, although our sample points to a sub-population of stars with slower initial rotations than commonly assumed (Figure~\ref{fig:comparison_with_models}). 

\subsection{Switching from Convective to Tachoclinal Dynamos}

The primary scientific question of the study is raised in the Introduction:  When during the complex changes in stellar structure of the l-PMS evolutionary phase does the magnetic dynamo change from the $\alpha^2$ distributed convective process to the solar-type $\alpha\Omega$ tachoclinal process?   The X-ray/rotation study of the l-PMS cluster h per by \citet{Argiroffi2016} gave evidence, based on a few slowly rotating stars, that the transition to an $\alpha\Omega$ dynamo was beginning for more massive stars at age $\simeq 13$~Myr.  Our $L_X-P_{rot}$ diagram for stars in a wider range of cluster ages from $\simeq 5-25$~Myr (Figure~\ref{fig:lx_p}) is consistent with this result but does not provide convincing independent support.  The distributions are also consistent with the hypothesis that X-ray luminosities have no relationship to rotational period in l-PMS stars. 

Our ZTF-based periods are limited to periods shorter than $\sim 10$ days where the stars are still in a magnetically saturated regime with Rossby number being below $\sim 0.1$.  Definitive results may require measuring rotation rates spectroscopically for mass-stratified samples of l-PMS covering a range of X-ray luminosities. 

However, this study together with \citet{Getman22} provide two other lines of evidence for dynamo changes.  First, Paper~I showed strong declines in X-ray luminosity in l-PMS stars. The decline is dramatic in higher-mass $1-3.5$~M$_\odot$ stars and slower in lower-mass $0.75-1$~M$_\odot$ stars. The $L_X \propto t^{-1.8}$ decline in $1-3.5$~M$_\odot$ l-PMS stars is consistent with that of young main sequence stars where surface activity is driven by the $\alpha\Omega$ dynamo. 

Second, a subtle effect is seen at the later stages of l-PMS evolution as stars approach the ZAMS.  There is clear evidence that the universal relation between $L_X$ and stellar magnetic flux $\Phi$ applies at all evolutionary phases from e-PMS through l-PMS, ZAMS and older main sequence stars.  Our empirical results support this with the correlation $L_X \propto A_{spot}^{1}$ (\S\ref{sec:spot_sizes}, Figure~\ref{fig:lx_aspot_temp_evol}, Table~\ref{tab:lx_aspot_slopes}). Further assuming convective dynamos are dominant, the $L_X \propto \Phi^1$ relation, where $\Phi$ is a theory based magnetic flux, holds for stars in the e-PMS and beginning of l-PMS stages (\S\ref{sec:magflux}, Figure~\ref{fig:fig_discussion}b,c). But the convective dynamo flux generation equation (\ref{eq:magnetic_strength}) fails to reveal the expected $L_X-\Phi$ correlation at later l-PMS stages (Figure~\ref{fig:fig_discussion}d).  This suggests that a tachoclinal $\alpha\Omega$ processes emerge as the dominant dynamo mechanism as stars move along the late part of Henyey track in the Hertzsprung-Russell diagram and approach the ZAMS.

\section{Acknowledgments}
We offer this study in memory of our late friend and colleague Leisa K. Townsley (Penn State). We are grateful to the anonymous referee for thoughtful and helpful comments that improved the manuscript. This project is supported by the {\it Chandra} grant GO9-20011X (K. Getman, Principal Investigator) and the {\it Chandra} ACIS Team contract SV4-74018 (G. Garmire \& E. Feigelson, Principal Investigators), issued by the {\it Chandra} X-ray Center, which is operated by the Smithsonian Astrophysical Observatory for and on behalf of NASA under contract NAS8-03060. The {\it Chandra} Guaranteed Time Observations (GTO) data used here were selected by the ACIS Instrument Principal Investigator, Gordon P. Garmire, of the Huntingdon Institute for X-ray Astronomy, LLC, which is under contract to the Smithsonian Astrophysical Observatory; contract SV2-82024. 

This research has also made use of the NASA/IPAC Infrared Science Archive, which is funded by the National Aeronautics and Space Administration and operated by the California Institute of Technology. Based on observations obtained with the Samuel Oschin Telescope 48-inch and the 60-inch Telescope at the Palomar Observatory as part of the Zwicky Transient Facility project. ZTF is supported by the National Science Foundation under Grants No. AST-1440341 and AST-2034437 and a collaboration including current partners Caltech, IPAC, the Weizmann Institute for Science, the Oskar Klein Center at Stockholm University, the University of Maryland, Deutsches Elektronen-Synchrotron and Humboldt University, the TANGO Consortium of Taiwan, the University of Wisconsin at Milwaukee, Trinity College Dublin, Lawrence Livermore National Laboratories, IN2P3, University of Warwick, Ruhr University Bochum, Northwestern University and former partners the University of Washington, Los Alamos National Laboratories, and Lawrence Berkeley National Laboratories. Operations are conducted by COO, IPAC, and UW. 

This work has made use of data from the European Space Agency (ESA) mission {\it Gaia} (\url{https://www.cosmos.esa.int/gaia}), processed by the {\it Gaia} Data Processing and Analysis Consortium (DPAC, \url{https://www.cosmos.esa.int/web/gaia/dpac/consortium}). Funding for the DPAC has been provided by national institutions, in particular the institutions participating in the {\it Gaia} Multilateral Agreement.

\vspace{5mm}
\facilities{CXO, Gaia, ZTF}

\software{R \citep{RCoreTeam20}}

\clearpage\newpage

\appendix

\section{NGC 2362 Properties}  \label{sec:appendix_NGC2362}

$Chandra$-ACIS-I 100~ks exposure of this cluster was analysed by our team as part of the 20-region Massive Young Star-forming complex study in Infrared and X-ray (MYStIX) survey \citep{Feigelson13}.  The X-ray study emerged with 491 X-ray selected cluster members tabulated by \citet{Broos13}.  Stellar X-ray luminosities and ages based on the $Age_{JX}$ chronometer \citep{Getman14a} are listed in Table~6 of Paper~I.

Here we re-examine the earlier distance and age estimates for the cluster using Gaia DR3 data \citep{GaiaMission2016,GaiaDR32022} and methods discussed in Paper~I. Four hundred forty six {\it Chandra} cluster members from \citet{Broos13} have Gaia DR3 counterparts within $1\arcsec$ matching radius.  The median distance from the Sun (with 68\% bootstrap error) based on 186 stars with Gaia re-normalized unit weight errors (RUWE) values $<1.4$ and parallax accuracy $\sigma_{\omega} < 0.1$~mas, are $1301 \pm 9$~pc. This is consistent with, and more accurate than, the previous Gaia DR2-based distance of $1332 \pm 70$~pc \citep{Kuhn19}. 

Fitting redden theoretical PARSEC 1.2S isochrones \citep{Bressan12,Chen14} to the Gaia DR3 color-magnitude distribution of 351 {\it Chandra} members with the error on the Gaia $G_{BP} - G_{RP}$ color of  $<0.1$~mag, gives a cluster extinction and age of $A_V = 0.3$~mag and $t=5$~Myr, respectively.  The extinction coefficients from \citet{Luhman2020} were used to redden the theoretical PARSEC isochrones. This isochronal fit is shown in the Gaia color-magnitude diagram in Appendix~\ref{sec:appendix_m_p}.  Following Paper~I, effective temperatures, bolometric luminosities, and masses for individual cluster members are obtained through a comparison of the observed $G$-band absolute magnitudes with theoretical predictions for the best-fit ($5$~Myr) reddened PARSEC 1.2S isochrone.  

Ninety-five X-ray cluster members are located in the central part of the cluster covered by the $\it Chandra$ observation; have Gaia DR3-based masses; and have rotation periods measured by \citet{Irwin2008} as part of the Monitor project.  The cluster was observed with a 6~minute cadence in the $i$ band for several hours on 8 nearly-contiguous nights with the CTIO Blanco 4-m telescope. Typical photometric precision is $2-3$~mmag at $i=15$~mag and $5-6$~mmag at $i=18$~mag. Rotational periods are obtained from least-squares fits to sinusoidal functions, a threshold on the reduced chi-squared value, and visual vetting. Reliability is established with simulations.   The last 95 records in Table~\ref{tab:p_lx_ztf_and_ngc2362} here present these stars and rotational periods.

The X-ray luminosities for NGC~2362 members are calculated using the XPHOT procedure \citep{Getman10}.  This requires the presence of both soft ($<2$~keV) and hard ($>2$~keV) X-ray photons, but many NGC~2362 members lack hard X-ray photons. Table~6 of \citet{Getman22} provides XPHOT-based intrinsic X-ray luminosities ($L_X$) for only 207 (out of 491) brighter X-ray members of NGC~2362. Using the 207 bright X-ray stars, we calculate a single apparent-to-intrinsic X-ray flux conversion factor by averaging the ratios of $L_X$s to apparent X-ray photometric fluxes $Photon\_Flux\_t$, given in Table~2 of \citet{Broos13}. This factor is applied to the apparent {\it Chandra} fluxes of the remaining weaker stars to determine their intrinsic X-ray luminosities. For all 95 NGC~2362 X-ray members with available rotation periods from \citet{Irwin2008}, their inferred intrinsic X-ray luminosities are listed in Table~\ref{tab:p_lx_ztf_and_ngc2362}. 

No additional non-X-ray Gaia members were added to this NGC~2362 stellar sample, because the X-ray sample by itself is already complete down to $M=0.6$~$M_{\odot}$, a low mass limit of interest in this X-ray-rotation study, thanks to the relatively long {\it Chandra} observation exposure time for this unobscured cluster.  Thus all X-ray upper limit records in Column 11 of Table~\ref{tab:p_lx_ztf_and_ngc2362} for this cluster are marked as ``...'' (no data).

\section{Gaia-DR3 Masses For Previously Published Stars With Rotation Periods} \label{sec:appendix_m_p}

\begin{figure*}[ht!]
\includegraphics[width=\textwidth]{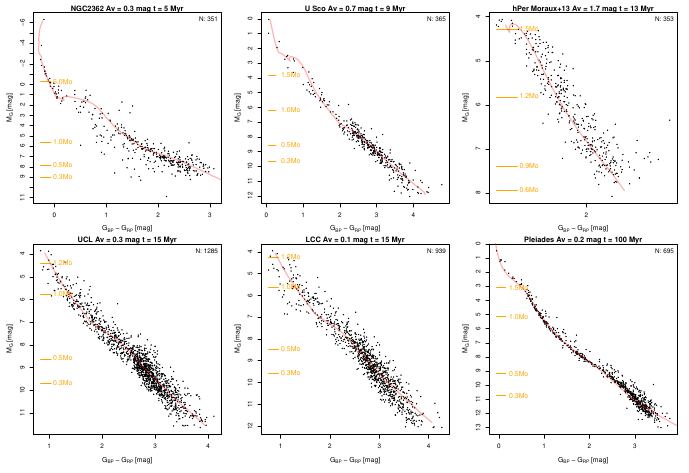}
\caption{Gaia-DR3 color-magnitude diagrams for stars with previously published rotation periods. For NGC~2362, the shown sample is the MYStIX X-ray sample discussed in \S\ref{sec:appendix_NGC2362}. For the remaining clusters, the shown stars are from Table~\ref{tab:gaia_masses_previous_stars}.  The red curves represent the best-fit reddened PARSEC 1.2S evolutionary models. Figure titles list the inferred best fit cluster extinction and age. Corresponding stellar masses are indicated by orange markers. Figure legends provide numbers of plotted stars.}  \label{fig:cmds_combined}
\end{figure*}

This section presents Table~\ref{tab:gaia_masses_previous_stars}, listing Gaia-based masses for stars most of which lack {\it Chandra}-X-ray data but have previously published rotation periods, as well as Figure~\ref{fig:cmds_combined} showing Gaia-DR3 color-magnitude diagrams for such stars.

\section{Updated HRD Phases for l-PMS Stars from Paper~I} \label{sec:appendix_hrd_phase}

Table~4 in Paper~I lists the properties of 7924 {\it Chandra}-Gaia stellar members of 10 open clusters. The Column~11 of that table gives an integer flag indicating the evolutionary status of the star on the Hertzsprung-Russell diagram based on the predictions of the PARSEC 1.2S evolutionary models. Original HRD statuses in the PARSEC 1.2S mass-track files of \citet{Chen14} are given as floating-point real numbers, with the integer parts indicating critical evolutionary points and the fractional parts indicating fractional time durations between the critical points. The critical points for young stars include: $=1$ - the onset of Hayashi track; $=2$ - onset of early Henyey track; $=3$ - onset of late Henyey track (when $L_{bol}$ reaches maximum on Henyey track); $=4$ - onset of ZAMS; $=5$ - onset of phase when hydrogen burning is fully active. According to \citet{Chen14}, fractional parts are ``proportional to the fractional time duration between the current point and the beginning of that phase'', as $frac = (t-t_{beg,i})/(t_{end,i}-t_{beg,i})$ and $i=\{1,2,3,4,5\}$ for young stars. Table~\ref{tab:new_hrd} provides more accurate (including the fractional parts) HRD statuses for the 7924 l-PMS stars of Paper~I. These new stellar HRD statuses are further employed in Figures \ref{fig:logp_logm} and \ref{fig:fig_discussion}.

\begin{deluxetable*}{cccccccccccc}
\tabletypesize{\footnotesize}
\tablecaption{Gaia-DR3 Masses For Stars With Previously Published Rotation Periods \label{tab:gaia_masses_previous_stars}}
\tablewidth{0pt}
\tablehead{
\colhead{Cluster} & \colhead{R.A.} &
\colhead{Decl.} & \colhead{$G$} & \colhead{$\sigma_{G}$} & \colhead{$B-R$} & \colhead{$\sigma_{BR}$} & \colhead{$\log(T_{eff})$} &
\colhead{$M$} & \colhead{$\log(L_{bol})$} & \colhead{$P_{rot}$} & \colhead{$Ampl$}\\
\colhead{} & \colhead{deg} &  \colhead{deg} & \colhead{mag} & \colhead{mag} & \colhead{mag} & \colhead{mag} & \colhead{K} & \colhead{$M_{\odot}$} & \colhead{$L_{\odot}$} & \colhead{day} & \colhead{mag}\\
\colhead{(1)} & \colhead{(2)} & \colhead{(3)} & \colhead{(4)} & \colhead{(5)} & \colhead{(6)} & \colhead{(7)} & \colhead{(8)} & \colhead{(9)} & \colhead{(10)} & \colhead{(11)} & \colhead{(12)}
}
\startdata
NGC2362 & 109.395103 & -25.168670 & 16.816 & 0.003 & 1.379 & 0.007 & 3.62 & 0.79 & -0.38 & 7.34 & 0.010\\
NGC2362 & 109.412882 & -25.192423 & 16.510 & 0.003 & 1.396 & 0.007 & 3.63 & 0.85 & -0.29 & 7.77 & 0.006\\
NGC2362 & 109.500341 & -25.137874 & 19.014 & 0.003 & 3.090 & 0.065 & 3.51 & 0.39 & -0.93 & 1.14 & 0.030\\
NGC2362 & 109.518947 & -25.240841 & 18.059 & 0.003 & 2.118 & 0.020 & 3.55 & 0.60 & -0.69 & 9.34 & 0.018\\
NGC2362 & 109.554830 & -25.236658 & 20.009 & 0.004 & 2.270 & 0.093 & 3.47 & 0.23 & -1.23 & 10.80 & 0.050\\
NGC2362 & 109.564310 & -25.247717 & 18.288 & 0.003 & 2.496 & 0.032 & 3.54 & 0.49 & -0.76 & 5.46 & 0.040\\
NGC2362 & 109.596027 & -25.194042 & 17.907 & 0.003 & 1.887 & 0.014 & 3.56 & 0.63 & -0.64 & 18.17 & 0.010\\
NGC2362 & 109.592932 & -25.261247 & 17.210 & 0.006 & 1.706 & 0.040 & 3.59 & 0.70 & -0.48 & 2.80 & 0.194\\
NGC2362 & 109.601028 & -25.141686 & 16.454 & 0.007 & 1.672 & 0.030 & 3.64 & 0.94 & -0.26 & 12.02 & 0.052\\
NGC2362 & 109.638649 & -25.191095 & 18.971 & 0.003 & 2.805 & 0.061 & 3.51 & 0.43 & -0.93 & 3.61 & 0.022\\
\enddata 
\tablecomments{This table is available in its entirety (3879 stars with Gaia properties and rotation periods) in the machine-readable form in the online journal. Gaia-based stellar properties (effective temperatures, masses, and bolometric luminosities) are estimated in the current paper. Rotation periods are from \citet[][NGC~2362]{Irwin2008}, \citet[][Upper Sco]{Rebull2018}, \citet[][NGC~869]{Moraux2013}, \citet[][UCL \& LCC]{Rebull2022}, and \citet[][Pleiades]{Rebull2016}. For NGC~2362 and NGC~869, there is overlap between the current sample and the X-ray-rotation sample given in Table~\ref{tab:p_lx_ztf_and_ngc2362}. Column 1: Cluster name. Columns 2-3: Gaia DR3 right ascension and declination (in decimal degrees) for epoch J2000.0. Columns 4-7: Gaia DR3 magnitude and color and their uncertainties. Only stars with $\sigma_{BR} < 0.1$~mag are included in this table and corresponding Figure~\ref{fig:cmds_combined}. Columns 8-10: Stellar effective temperature, mass, and bolometric luminosity derived from the Gaia color-magnitude diagrams. Columns 11-12: Rotation periods and peak-to-peak amplitudes from \citet{Irwin2008,Moraux2013,Rebull2016,Rebull2018,Rebull2022}. The amplitudes are only available for NGC~2362 and NGC~869.}
\end{deluxetable*}

\begin{deluxetable*}{cccc}
\tabletypesize{\footnotesize}
\tablecaption{New HRD Statuses For l-PMS Stars From Paper~I \label{tab:new_hrd}}
\tablewidth{0pt}
\tablehead{
\colhead{Cluster} & \colhead{R.A.} &
\colhead{Decl.} & \colhead{$ph_{HRD}$}\\
\colhead{} & \colhead{deg} &  \colhead{deg} & \colhead{}\\
\colhead{(1)} & \colhead{(2)} & \colhead{(3)} & \colhead{(4)}
}
\startdata
NGC1502 & 62.257371 & 62.385897 & 1.11\\
NGC1502 & 61.944486 & 62.153143 & 5.00\\
NGC1502 & 61.577576 & 62.266908 & 1.25\\
NGC1502 & 61.904523 & 62.509605 & 1.20\\
NGC1502 & 62.374648 & 62.351225 & 1.11\\
NGC1502 & 61.665573 & 62.324773 & 2.05\\
NGC1502 & 61.679861 & 62.340361 & 2.23\\
NGC1502 & 61.758327 & 62.315094 & 1.71\\
NGC1502 & 61.764918 & 62.335727 & 1.11\\
NGC1502 & 61.769816 & 62.227697 & 2.44\\
\enddata 
\tablecomments{This table is available in its entirety (7924 stars) in the machine-readable form in the online journal. Additional stellar properties for these l-PMS stars including the X-ray and bolometric luminosities, stellar masses, and effective temperatures can be found in Table~4 of Paper~I. Column 1: Cluster name. Columns 2-3: Gaia DR3 right ascension and declination (in decimal degrees) for epoch J2000.0. Columns 4: Updated status of the stellar evolutionary phase. See text for details.}
\end{deluxetable*}

\clearpage\newpage

\bibliography{my_bibliography}{}
\bibliographystyle{aasjournal}

\end{document}